\documentstyle [11pt]{article}
\newcommand{\f}{\left( \frac{i \hbar c}{e} \right)}
\newcommand{\crr}{\left(\frac{-ie}{\hbar c} \right)}
\newcommand{\r}{(\frac{e}{c})}
\begin{document}
\title{ON THE HAMILTONIAN APPROACH AND PATH INTEGRATION
 FOR A POINT PARTICLE  MINIMALLY COUPLED TO ELECTROMAGNETISM}
\author{Kostas Skenderis\thanks{ e-mail: kostas@max.physics.sunysb.edu }
\ and Peter van Nieuwenhuizen\thanks{ e-mail:
vannieu@max.physics.sunysb.edu } \\
                                  \\
Institute for Theoretical Physics, \\
State University of New York at Stony Brook, \\
Stony Brook, NY 11794-3840 }
\maketitle
\begin{abstract}
We {\em derive} the exact configuration space path integral, together with
the way how to evaluate it, from the Hamiltonian approach for any quantum
mechanical system in flat spacetime whose Hamiltonian has at most two
momentum operators. Starting from a given, covariant or non-covariant,
Hamiltonian, we go from the time-discretized path integral to the continuum
path integral by introducing Fourier modes. We prove that the limit
$N \rightarrow \infty$ for the terms in the perturbation expansion
(``Feynman graphs'') exists, by demonstrating that the series involved
are uniformly convergent.
{\em All} terms in the expansion of the exponent in
$<x| \exp (- \Delta t \hat{H} / \hbar) |y>$
contribute to the propagator (even at order $\Delta t$!). However,
in the time-discretized path integral the only effect of the terms with
$\hat{H}^2$ and higher is to cancel terms  which naively seem to vanish
for $N \rightarrow \infty$, but, in fact, are nonvanishing. The final
result is that the naive correspondence between the Hamiltonian and the
Lagrangian approach is correct, after all. We explicitly work through
the example of a point particle coupled to electromagnetism. We compute
the propagator to order $(\Delta t)^2$ both with the Hamiltonian and
the path integral approach and find agreement.
\end{abstract}
\pagebreak

\section{Introduction.}

Path integrals are often first written down in a symbolic way as an
integral over paths of the exponent of an action,
and then defined by some time-discretization. Of course, there are many
ways in which to implement time-discretization.
In some instances, rules have been discovered which lead to desirable
answers for the path integral. A well-known
example is the mid-point rule for the interaction
$\int dt A_j(x) \dot{x}^j$ of a point particle coupled to an
electromagnetic potential. This rule leads to gauge invariance
of the terms proportional to $\Delta t$ in the
propagator\cite{schul} but it breaks gauge invariance in the terms of
order $(\Delta t)^2$. The terms of higher order in $\Delta t$ are needed
for the evaluation of anomalies (see below).
In general, starting from a continuum path integral, there is no preferred
way to discretize it. One might take the point of view that different
discretizations simply correspond to different theories.

In this article we take a different point of view. We take the Hamiltonian
formalism as starting point, and shall
{\em deduce} both the action $S_{\rm config}$ to be used in the
configuration space path integral, and the way this path integral
should be evaluated (``the measure''). We mean by the expressions
``to be used'' and ``should be'' that in this way the path integral
formalism exactly reproduces the propagator  of the Hamiltonian formalism.
Of course, in the Hamiltonian $\hat{H}(\hat{x}^i,\hat{p}_j)$ there is a
priori a corresponding ambiguity in the ordering of the operators
$\hat{x}^i$ and $\hat{p}_j$. However, in several examples, the Hamiltonian
of a quantum mechanical model is, in fact, the regulator of the Jacobians
for symmetry transformations of a corresponding {\em field} theory, and
these regulators are uniquely fixed by requiring certain symmetries of the
field theory to be maintained at the quantum level\cite{A_W,diaz,fio,vanfio}.
For example, in \cite{diaz} regulators are constructed which maintain Weyl
(local scale) invariance but as a
consequence break Einstein (general coordinate) invariance.
Thus: the field theory and the choice of which symmetries are free of
anomalies fixes the regulator, the regulator is the Hamiltonian of a
corresponding quantum mechanical model, and the operator ordering of this
Hamiltonian is thus fixed.
For these reason we consider Hamiltonians of the form
$ \hat{H}=\hat{p}^2 + a^i(\hat{x}) \hat{p}_i + b(\hat{x})$
whose operator ordering is fixed in this way but whose coefficents
$a^i(\hat{x})$ and $b(\hat{x})$ are not restricted
except that we assume that they are regular functions; they may correspond
to  covariant or non-covariant Hamiltonians. The results of this paper
{\em prove} which path integral (including, of course, the way how to
evaluate it) corresponds to which regulator (Hamiltonian). For chiral
anomalies\cite{A_W} this precise correspondence was not needed due to
their topological nature, but for trace anomalies\cite{fio,vanfio} and
other anomalies of non-topological nature, the precise correspondence is
crucial.

Having obtained the 1-1 correspondence, it is then also possible to start
with a particular action in the path integral (the latter to be evaluated
as derived below) and to find the corresponding Hamiltonian operator. This
will usually be the case when one is dealing with quantum field theories.
For example, when one is dealing with renormalizable field theories or when
the theory has certain symmetries the action may be known, and this will fix
the operator ordering and the terms in the Hamiltonian.

In the Hamiltonian approach the propagator is defined by
\begin{equation}
<x,t_2|y,t_1> = <x| \exp (-\frac{\Delta t}{\hbar} \hat{H}|y>,
\; \Delta t=t_2-t_1,
\end{equation}
and evaluated, following Feynman, by inserting a complete set of
momentum eigenstates
\begin{equation}
<x,t_2|y,t_1> = \int dp <x| \exp (-\frac{\Delta t}{\hbar} \hat{H}|p><p|y>.
\end{equation}
Expanding the exponent, and moving in each term
$(-\Delta t \hat{H} / \hbar)^n /n!$ the $\hat{x}$
operators to the left and the $\hat{p}$ operators to the right,
one obtains an {\em unambiguous} answer for the propagator.
No regularization is needed. However, one must keep track of the
commutators between $\hat{x}^i$ and $\hat{p}_j$. It is often assumed
that it is sufficient to expand the exponent only to first order in
$\Delta t$, and to reexpontiate the result
\begin{eqnarray}
<x| \exp (-\frac{\Delta t}{\hbar} \hat{H}|p>
& = &\exp [-\frac{\Delta t}{\hbar} h(x,p)]<x|p>
\; \mbox{(false!)} \label{la} \\
<x|\hat{H}|p>& \equiv &h(x,p)<x|p>. \label{linapp}
\end{eqnarray}
This is incorrect for Hamiltonians with derivative coupling:
for nonlinear sigma models where the $\hat{p}^2$ term is multiplied
by a function of $\hat{x}$ (``the metric'')\cite{graham,book,bas}
or for Hamiltonians with a term
$A(\hat{x}) \cdot \hat{p}$. We shall consider the Hamiltonian
\begin{equation}
\hat{H} = \frac{1}{2} \Big( \hat{p}_i - (\frac{e}{c}) A_i(\hat{x}) \Big)
\Big( \hat{p}^i - (\frac{e}{c}) A^i(\hat{x}) \Big) + V(\hat{x}),
\end{equation}
for arbitrary but nonsingular $A_i(\hat{x})$ and $V(\hat{x})$ which is
obviously the most general Hamiltonian of the form
$ \hat{H}=\hat{p}^2 + a^i(\hat{x}) \hat{p}_i + b(\hat{x})$,
and show that there are terms proportional to $\Delta t$ in the
propagator which are due to commutators between
$\hat{p}_i$ and $A_j(\hat{x})$. In fact, {\em all} terms in the
expansion of the exponential give such
contributions\cite{graham,book,bas}!

Because the commutators $[\hat{p}^i, \hat{x}_j]=-i \hbar \delta^i_j$
are proportional to $\hbar$, the propagator becomes a series in $\hbar$,
$\Delta t$ and $(x-y)^i$ with coefficients which are functions of $x$.
When we use the term ``of order $(\Delta t)^k$'' we mean all terms which
differ from the leading term by a factor $(\Delta t)^k$, counting
$(x-y)^i$ as $(\Delta t)^{1/2}$.
The terms of order $\hbar$ w.r.t. the classical result correspond to
one-loop corrections in the path integral approach, and can be written
in terms of the classical action as the Van Vleck
determinant\cite{vleck,witt}. Terms of higher order in $\hbar$ in
the propagator can be computed straightforwardly (though tediously)
in the Hamiltonian approach, again without need to specify a regularization.
This indicates that the details of the path integral should follow
straightforwardly from the Hamiltonian starting point. In particular it
should not be necessary to fix a free constant in the overall normalization
of the path integral by hand, for example by dividing by the path integral
for a free particle.

One begins by defining the path integral as
\begin{equation}
<x,0|y,-T> = \lim_{N \rightarrow \infty} \int
\Big[ \prod_{\alpha=1}^{N-1} dx_\alpha \Big]
\Big[ \prod_{\alpha=1}^N <x_{\alpha-1},t_{\alpha-1}|x_\alpha,t_\alpha> \Big],
\end{equation}
where $x_0 = x$ and $x_N = y$. This particular time-discretization
follows from the Hamiltonian approach; it is due to the operator identity
\begin{equation}
\exp ( - \frac{T}{\hbar} \hat{H} ) =
\Big( \exp ( - \frac{T/N}{\hbar} \hat{H} ) \Big)^N.
\end{equation}
The main result of this paper is a proof that the $N \rightarrow \infty$
limit exists, and defines a continuum action $S_{\rm config}$ and an
unambiguous and simple way to evaluate the path integral perturbatively.

We begin by  splitting $x_\alpha$ into a background part $z_\alpha$ and a
quantum part $\xi_\alpha$.
We shall also decompose the time-discretized action $S$ into a part $S_0$
which yields the propagator on the world line, and the rest which yields the
interaction terms $S_{\rm int}$.  The $z_\alpha$ satisfy the equation of
motion of $S_0$ and the boundary conditions $z_\alpha=y$ at $\alpha=N$ and
$z_\alpha=x$ at $\alpha=0$, so that $\xi_\alpha=0$ both at $\alpha=0$ and
at $\alpha=N$. Since $S_0$ is not equal to $S$, there are  terms linear in
$\xi_\alpha$ in the expansion of
$S(z+\xi,\dot{z}+\dot{\xi})$. Notice that the time-discretized action $S$
has {\em not} been obtained by some ad-hoc rule, but rather it is determined
from the Hamiltonian approach.

The final result for the path integral should not depend on the choice of
$S_0$. We choose $S_0$ as the action of a free particle because that leads
to simple perturbation theory, but other choices of $S_0$ should lead to
the same final result although the Feynman rules for the perturbative
expansion of the path integral will be different.

One now expands $\xi_\alpha$ into eigenfunctions of $S_0$, i.e., in terms
of trigonometric functions
\begin{equation}
\xi_\alpha = \sum_{k=1}^{N-1} y^k \sin \alpha k \pi / N,
\; (\alpha=1,\ldots,N-1).
\end{equation}
Changing integration variables from $\xi_\alpha$ to $y^k$, the Jacobian is
essentially unity, while $S_0$ is quadratic and diagonal in these $y^k$.
Rescaling these $y^k$ such that the kinetic term in terms of the rescaled
variables $v^k$ becomes the one of the continuum theory, the Jacobian of
this rescaling leads to a non-trivial factor in the measure. At this point,
the path integral has the generic form
\begin{equation}
\int d\mu \exp [-\frac{1}{\hbar} (S_0 + S_{\rm int}(N))],
\end{equation}
where the measure $\mu$ and the kinetic term $S_0$ are already in the form
of the continuum theory, but the interaction $S_{\rm int}$ still depend
on $N$.

By the term ``continuum theory'' we mean the path integral with \newline
(i) the classical Lagrangian
$L=\frac{1}{2} \dot{x}^2 - i(\frac{e}{c}) \dot{x}^i A_i + V$, \newline
(ii) the expansion $x(t) = z(t) + \sum_{k=1}^{\infty} v^k \sin k \pi t/T$,
where $z(t)$ is a solution of the equation of motion $\ddot{z}=0$ with
the boundary conditions $z(0)=x$ and $z(T)=y$, and   \newline
(iii) the measure which normalizes the Gaussian integration with $S_0$
over the modes $v^k$ to
$(2 \pi \hbar T)^{-1/2}$ (not to unity because there is always one more
intermediate set $|p><p|$ in Feynman's approach than intermediate sets
$|x><x|$. The remaining factor $(2 \pi \hbar T)^{-1/2}$ combines with
classical part $\exp \big(-(x-y)^2 /2 \hbar T\big)$ to yield a
representation of $\delta(x-y)$ for small $T$).

One must then show that the limit $N \rightarrow \infty$ in $S_{\rm int}$
yields the interaction of the continuum theory. This is a well-known
complicated problem, but we shall present here a totally elementary proof
which uses only trigonometric relations such as
$2 \sin \alpha \sin \beta = \cos (\alpha-\beta) - \cos (\alpha+\beta)$
and the fact that the infinite series we encounter are uniformly convergent
as functions of $N$. This property allows us to take the limit
$N \rightarrow \infty$ before the summations are performed.
For the Hamiltonians of the form $T(p)+V(x)$ such a  proof is quite simple,
but for non-vanishing vector potential $A(x)$, we need rather laborious
algebra.

The result is surprisingly simple. All terms in the propagator which in
the Hamiltonian approach are due to commutators, are only needed to make
sure that in the limit $N \rightarrow \infty$ one obtains the classical
action. To be more explicit, consider the discretized action in
(\ref{inte}).
The last three lines vanish in the naive limit
$N \rightarrow \infty$ (the limit $N \rightarrow \infty$ for fixed mode
index $k$) since they have extra factors $1/N$ w.r.t. the two lines
above. These latter two lines naively yield the term
$\int dt A_j(x) \dot{x}^j$ in the
classical action because $1/N$ becomes $dt$ and
$(\xi_{\alpha-1} - \xi_\alpha)$ becomes $\dot{\xi} dt$.
The claim is that if one does not take the naive limit but
carefully evaluates the sums, then the non-naive terms in the first
two lines cancel all of the last three lines in (\ref{inte}).
To discuss in more detail what we mean by the naive limit
$N \rightarrow \infty$ consider the interaction terms
\begin{equation}
S_{\rm int}= \frac{1}{N} \sum_{k,l=1}^{N-1} v^k v^l \lambda(k) \lambda(l)
\sum_{\alpha=1}^{N-1} \sin \alpha k \pi /N \sin \alpha l \pi /N,
\end{equation}
where $\lambda(k)=k \pi [2N^2(1-\cos k \pi /N)]^{-1/2}$.
For fixed $k$ and $N$ tending to infinity one finds,
$\lambda(k)=1$ while for $k \sim N$ tending to infinity one has
 $\lambda(k)=\pi/2$. One {\em may} take the limit
$N \rightarrow \infty$ in $S_{\rm naive}(N)$ for given fixed
$k$ and $l$, because the error thus committed  cancels against
the terms in $S(N) - S_{\rm naive}(N)$. Here $S_{\rm naive}(N)$
is the time-discretized action we would have obtained, if we had
ignored the terms coming from commutators in the Hamiltonian approach.

The non-trivial measure factorizes into a factor for each mode $v^k$.
One can then easily compute propagators
$\langle v^k v^l \rangle$ and Feynman graphs in terms of modes.
One can also use the quantum ``fields'' $\xi(\tau)$ and find that
$\langle \xi(\tau_1) \xi(\tau_2) \rangle$ is the expected world
line propagator (the inverse of
$\partial^2 / \partial \tau^2$
with the correct boundary conditions). However, the mode
representation is to be preffered because mode cut-off is the
natural regularization scheme\cite{fio,vanfio}. Actually, all
one-loop diagrams we evaluate are already finite by themselves
since the divergences of the tadpole graphs are put to zero by
mode regularization. Although we do not consider here curved space,
we mention for completeness that in curved spacetime there are
extra ``ghosts'' obtained by exponentiating a factor
$(\det g_{ij})^{1/2}$ in the measure and that with these ghosts
all loop calculations become finite if one uses mode
regularization\cite{fio,vanfio,bas}.

{}From our point of view, the ambiguities often encountered in the
definition of path integrals are due to starting ``halfway''.
Starting from the beginning, which means for us starting with the
Hamiltonian approach, no ambiguities result and one {\em derives}
the action to be used in the path integral. The result is the 1-1
correspondence
\begin{eqnarray}
\hat{H} = \frac{1}{2} \; (\hat{p}_i - (\frac{e}{c}) \hat{A}_i)
&\delta^{ij}&(\hat{p}_j - (\frac{e}{c}) \hat{A}_j) \;
+ \; \hat{V} \nonumber \\
&\Updownarrow&  \\ \label{cor}
S_{\rm config} = \int_{-T}^0 dt \big[ \frac{1}{2} \;
\delta_{ij} \dot{x}^i \dot{x}^j\!
&\!-&\!i (\frac{e}{c}) \dot{x}^i A_i \; + \; V \big], \nonumber
\end{eqnarray}
and the path integral is perturbatively evaluated by computing
Feynman graphs with given propagators and vertices.

In section 2 we discuss the Hamiltonian approach for a point
particle coupled to electromagnetism. Although we only need
the propagator to order $\Delta t$ in order to construct the
corresponding path integral, we evaluate it to order
$(\Delta t)^2$ in order to compare later with a similar result
obtained from the path integral. A useful check is that it
factorizes into a classical part and a Van Vleck determinant.
In section 3, the path integral is cast into a form where the
only $N$-dependence resides in the interaction terms $S_{\rm int}$.
In section 4, we discuss the limit $N \rightarrow \infty$ in
$S_{\rm int}$. We organize the discussion by giving six examples
which cover all possible cases one encounters in a perturbative
evaluation of the path integral.
In section 5 we evaluate as a check the path integral to order
$(\Delta t)^2$ at the one-loop level.
Here we discuss how to evaluate the continuum path integrals in
general in perturbation theory.
The result agrees with the one obtained in section 2 from the
Hamiltonian approach.
In section 6 we note that our work straightforwardly extends
to field theories with derivative coupling like Yang-Mills theory.
We discuss how our work might be extended to curved spacetime,
and also to phase space path integrals.

\section{Hamiltonian operator approach.}

We wish to evaluate the propagator in Euclidean space
\begin{equation} <x|\exp(- \Delta t \hat{H} / \hbar)|y>,  \end{equation}
where
\begin{equation}
\hat{H}=\frac{1}{2} \left( \hat{p}_{i}- ( \frac{e}{c} ) A_{i}(\hat{x}) \right)
\left( \hat{p}^{i}- ( \frac{e}{c} )A^{i}(\hat{x}) \right), \;
i=1, \ldots ,n. \label{ham}
\end{equation}
We do not add a term $V(x)$ since the analysis for this term is the same
as for the term $(A_i(x))^2$.
Indices are raised and lowered by $\delta_{ij}$, so for notational
simplicity we write all indices down.
We shall only use the commutation relation $[ \hat{p}_i ,\hat{x}_j ]
= \frac{\hbar}{i} \delta_{ij}$ and
$\hat{p}_i |p> = p_i |p>$ on momentum eigenstates $|p>$.
We rewrite the Hamiltonian as
\begin{equation}
\hat{H}= \hat{\alpha} - ( \frac{e}{c} ) \hat{\beta}
+ ( \frac{e}{c} )^2 \hat{\gamma}, \label{ham1}
\end{equation}
where \[ \hat{\alpha} = \frac{1}{2}\hat{p}^{2},~\;  \hat{\beta}=\hat{A}
\cdot \hat{p}, ~
\; \hat{ \gamma } = \frac{1}{2} \left( i ( \frac{ \hbar c}{ e } )
\partial \cdot \hat{A} + \hat{A}^{2} \right). \]
Following Feynman we insert a complete set of $|p>$ states
\begin{equation}
<x|\exp(- \Delta t \hat{H} / \hbar)|y>=
\int d^{n}p<x| \exp(- \Delta t \hat{H}/\hbar)|p><p|y>.
\end{equation}
We expand the exponential and define
\begin{equation}
<x|(\hat{H})^{k}|p>=\sum_{l=0}^{2k}B_{l}^{k}(x)p^{l}<x|p>,  \label{expa}
\end{equation}
where $B_{l}^{k}(x)p^{l}$ is a polynomial of degree l in p's, and
\begin{equation}
<x|p>= (2\pi\hbar)^{-n/2} \exp(\frac{i}{\hbar}x \cdot p).
\end{equation}

After rescaling the momenta as $p=\sqrt{ \hbar / \Delta t } \, q $ we have
\begin{eqnarray}
<x|\exp(- \Delta t \hat{H} / \hbar)|y>=
(4\pi^{2}\hbar\Delta t)^{-n/2}\int d^{n}q
\exp( i \frac{q \cdot (x-y)}{ \sqrt{\hbar \Delta t} })  \nonumber \\
\sum_{k=0}^{\infty}\frac{(-1)^{k}}{k!}
\sum_{l=0}^{2k}B_{l}^{k}(x)q^{l}(\frac{\Delta t}{\hbar})^{k-l/2}.
\end{eqnarray}
The leading term comes from summing all the terms with $ l= 2k $ and has
the simple form $ \exp ( - \frac{1}{2} q^2 )$. For this reason we
introduced the $q$ variable. It follows that only a finite number of
B's need to be calculated in order to obtain the result up to desired
order in $\Delta t$. In particular, the result up to and including
$(\Delta t)^{2}$ needs the first five B's ($l=2k$ through $l=2k-4$).
A detailed discussion of the combinatorics is given in \cite{bas}.
Here we merely give our result.
\begin{equation}
B_{2k}^{k}(x) q^{2k}  =  \alpha^{k},
\end{equation}
\begin{equation}
B_{2k-1}^{k}(x) q^{2k-1}  =  - ( \frac{e}{c} ) k \alpha^{k-1} \beta,
\end{equation}
\begin{eqnarray}
B_{2k-2}^{k}(x) q^{2k-2} & = &  \r^{2} k \alpha^{k-1} \gamma +
\nonumber  \\
 &\ &  \r^{2} \left(
\begin{array}{c} k \\ 2 \end{array}
\right)
\alpha^{k-2} \Biggl[ \f q_{i}(\partial_{i} \beta)+\beta^{2} \Biggr],
\label{exam}
\end{eqnarray}
\begin{eqnarray}
B_{2k-3}^{k}(x) q^{2k-3} = &-& \r^{3}
\left(
\begin{array}{c} k \\ 2 \end{array}
\right)
\alpha^{k-2} \Biggl[ \frac{1}{2} \f^2 \partial^{2} \beta
+ \f q_{i} \partial_{i} \gamma \nonumber \\
& \  &  \hspace{0.8cm} + 2 \beta \gamma
+ \f A_{i}( \partial_{i} A_{j})q_{j} \Biggr]  \nonumber \\
& - &  \r^{3}
\left(
\begin{array}{c} k \\ 3 \end{array}
\right)
\alpha^{k-3} \Biggl[ \f^2 q_{i} q_{j} \partial_{i} \partial_{j} \beta
\nonumber \\
 &\ &  \hspace{1cm} + 3 \f \beta q_{i} \partial_{i} \beta + \beta^{3} \Biggr],
\end{eqnarray}
\begin{eqnarray}
B_{2k-4}^{k}(x) q^{2k-4} & = & \r^{4}
\left(
\begin{array}{c} k \\ 2 \end{array}
\right)
\alpha^{k-2} \Biggl[ \frac{1}{2} \f^2 \partial^{2} \gamma
+ \f A_{i} \partial_{i} \gamma
+ \gamma^{2} \Biggr]  \nonumber \\
& + &  \r^{4}
\left(
\begin{array}{c} k \\ 3 \end{array}
\right)
\alpha^{k-3}
\Biggl[ \f^3 q_{i} ( \partial_{i} \partial^{2} \beta ) \nonumber \\
&\  & \hspace{0.8cm} + \f^2 [ q_{i} q_{j} ( \partial_{i} \partial_{j} \gamma )
+ (\partial_{i} \beta) (\partial_{i} \beta) \nonumber \\
&\  & \hspace{0.8cm} + \frac{3}{2}  \beta \partial^{2} \beta
+  q_{i} ( \partial_{i} A_{j} ) (\partial_{j} \beta)
+ 2 q_{i} A_{j} ( \partial_{i} \partial_{j} \beta ) ]\nonumber \\
&\  & \hspace{0.8cm} + \f [ 3 \beta q_{i} ( \partial_{i} \gamma )
+ 3  \gamma q_{i} ( \partial_{i} \beta )\nonumber \\
&\  & \hspace{0.8cm} + 3 A_{i} ( \partial_{i} \beta ) \beta ]
+ 3 \beta^{2} \gamma \Biggr] \nonumber \\
& + & \r^{4}
\left(
\begin{array}{c} k \\ 4\end{array}
\right)
\alpha^{k-4} \Biggl[ \f^3 q_{i} q_{j} q_{k}
( \partial_{i} \partial_{j} \partial_{k} \beta ) \nonumber \\
&\ & \hspace{0.8cm} + \f^2
[ 3 ( \partial_{i} \beta )( \partial_{j} \beta ) q_{i} q_{j}
+ 4 ( \partial_{i} \partial_{j} \beta ) \beta q_{i} q_{j} ] \nonumber \\
&\  & \hspace{0.8cm} +
6 \f \beta^{2}( \partial_{i} \beta ) q_{i} + \beta^{4} \Biggr].
\end{eqnarray}
The combinatorial factors
\[ \left(\begin{array}{c} k \\ s \end{array} \right) \]
indicate that only s out of k factors of $\hat{H}$ are involved in
yielding commutators. For example, the last term in (\ref{exam}) is
due to picking two factors $\beta$ and $k-2$ factors of $\alpha$ out
of $k$ factors $\hat{H}$. Clearly this can be done in
\[ \left(\begin{array}{c} k \\ 2 \end{array} \right) \]
ways and there are two powers of $q$ less than in the leading term.
Similarly, the one but last term in (\ref{exam}) is due to one commutator
of $\hat{\alpha}$ and $\hat{\beta}$.

Next we perform the summations over k which is easy and the Gaussian
integrals which are straightforward but tedious.
The result reads
\begin{eqnarray}
&\ & <x| \exp(- \Delta t \hat{H} / \hbar)|y> =
(2 \pi \hbar \Delta t )^{-n/2} \exp \big( - \frac{1}{2 \Delta t \hbar }
(x-y)^2 \big) \nonumber \\
&\ & \Biggl\{ 1- \crr A_i (x-y)_i \nonumber \\
&\ & \hspace{1cm} + \frac{1}{2} \Biggl[ \crr A_{i,j}
+ \crr^2 A_i A_j \Biggr] (x-y)_i (x-y)_j \nonumber \\
&\ & + \frac{1}{3!} \Biggl[ \crr A_{i,jk} +3 \crr^2 A_{i,j} A_k \nonumber \\
&\ & \hspace{1cm} + \crr^3 A_i A_j A_k \Biggr]
(x-y)_i (x-y)_j (x-y)_k \nonumber \\
&\ & + \frac{1}{4!} \Biggl[ \crr A_{i,jkl} +3 \crr^2 A_{i,j} A_{k,l}
+ 4 \crr^2 A_{i,jk} A_{l} \nonumber \\
&\ & \hspace{1cm} + 6 \crr^4 A_i A_j A_k A_l \Biggr]
(x-y)_i (x-y)_j (x-y)_k (x-y)_l \nonumber \\
&\ & + \frac{1}{4!} \frac{ \Delta t  }{ \hbar } \r^2
F_{ik} F_{kj} (x-y)_i (x-y)_j \nonumber \\
&\ & + \frac{i \Delta t }{12} \r \Bigl[ F_{ki,k} (x-y)_i
- \frac{1}{2} F_{ki,kj} (x-y)_i (x-y)_j \Bigr] \nonumber \\
&\ & - \frac{1}{12} \frac{ \Delta t }{ \hbar } \r^2  A_{i}
F_{kj,k} (x-y)_i (x-y)_j
- \frac{ (\Delta t)^2}{48} \r^2 F^2 \Biggr\}.  \label{res}
\end{eqnarray}

Before going on, we briefly compare this result with the incorrect result
which one would have obtained by the linear approximation mentioned in the
introduction and widely used. In the latter case we find instead of the the
terms in the curly brackets the following expression,
\begin{eqnarray}
&\ & <x| \exp(- \Delta t \hat{H} / \hbar)|y> =
(2 \pi \hbar \Delta t )^{-n/2} \exp \big( - \frac{1}{2 \Delta t \hbar }
(x-y)^2 \big) \nonumber \\
&\ & \Biggl\{ 1- \crr A_i (x-y)_i
+ \frac{1}{2} \crr^2 A_i A_j (x-y)_i (x-y)_j \nonumber \\
&\ & \hspace{1cm} - \frac{1}{2} i \Delta t \r \partial \cdot A \Biggr\}
\mbox{ (false).} \label{incor}
\end{eqnarray}
This result is obtained by replacing
$\int dp <x| \exp (- \Delta t \hat{H} / \hbar) |p> <p|y>$ by
$\int dp  \exp (- \Delta t\ h(x,p) / \hbar) <x|p> <p|y>$,
where $h(x,p)$ is defined in (\ref{linapp}).
The $p$-dependence in the exponent is coming from the $p^2$ term in
$\alpha$, a single $p$ in $\beta$ and the inner product
$p \cdot (x-y)$ from the plane waves.
Integration over $p$ yields (\ref{incor}) to order
$\Delta t$. To order $\Delta t$ we thus find the same number of terms in
both cases, but the term $\partial \cdot A$ is present only in the linear
approximation, whereas in the correct approach it is cancelled by a
commutator $ \big[ p^2, \beta \big] $. This commutator yields a term
$ p_i p_j \partial_i A_j $ whose integration over $p$ gives a
$\delta_{ij}$ term which cancel the $ \partial \cdot A $ term, and a term
with $ (x-y)_i (x-y)_j $ which is the term with $A_{i,j}$ in (\ref{res}).
The corrections due to commutators already show up at order $\Delta t$.
Clearly, the linear approximation gives already incorrect results for the
propagator at order $\Delta t$.

We expect the result in (\ref{res}) to contain a factor of
$\exp[- \frac{1}{\hbar} S_{cl} ]$, where $S_{cl}$ is the
classical action evaluated along a classical trajectory.
We claim that to order $(\Delta t)^2$ it reads
\begin{eqnarray}
S_{cl} & = & \frac{1}{2 \Delta t} (x-y)_i (x-y)_i - i \r \big\{ A_i (x-y)_i
- \frac{1}{2} A_{i,j} (x-y)_i (x-y)_j \nonumber \\
& + & \frac{1}{3!} A_{i,jk} (x-y)_i (x-y)_j (x-y)_k \nonumber \\
& - &  \frac{1}{4!} A_{i,jkl} (x-y)_i (x-y)_j (x-y)_k (x-y)_l \big\}
\nonumber \\
& - & \frac{ \Delta t}{24} \r^2 F_{ik} F_{kj} (x-y)_i (x-y)_j
+  0 ( (\Delta t)^{5/2} ). \label{action}
\end{eqnarray}
To obtain this result, we used that the classical Lagrangian corresponding
to (\ref{ham}) is given by
\begin{equation}
L= \frac{1}{2} \dot{x}_{i} \dot{x}^{i} -i \r \dot{x}^{i} A_{i},  \label{lag}
\end{equation}
where the factor `$i$' is due to our working in Euclidean space.
(The simplest way to see this is to note that one has the same
Hamiltonian in both the Minkowski and the Euclidean case, but one uses
$ \exp ( - i \Delta t \hat{H} / \hbar ) $ in the former and
$ \exp ( - \Delta t \hat{H} / \hbar ) $ in the latter case. In both cases
$<x|p>$ and $<p|y>$ are plane waves).
The dynamical equations of motion are
\begin{equation}
\ddot{x}_i = -i \r F_{ij} \dot{x}_j. \label{eqm}
\end{equation}
We then evaluated  $S_{cl}$ by expanding all fields around the endpoint $x$,
\begin{eqnarray}
S_{cl}=\int_{- \Delta t}^{0} L dt & = & \Delta t L(x)
- \frac{1}{2} (\Delta t)^{2} \frac{ dL(x) }{ dt } \nonumber \\
& + & \frac{1}{3!} (\Delta t)^{3} \frac{ d^{2} L(x) }{ dt^{2} }
- \frac{1}{4!} (\Delta t)^{4} \frac{ d^{3} L(x) }{ dt^{3} } + \cdots .
\label{exp}
\end{eqnarray}
We only need $x(t)$ and $\dot{x} (t)$ at $t=0$ since higher time derivatives
of $x(t)$ can be obtained by using the equations of motion (\ref{eqm}).
To obtain $\dot{x} (t=0)$ in terms of $x$ and $y$,
we  expand $x(-\Delta t)$ around $t=0$, and use
(\ref{eqm}). This yields a series in power of $\dot{x}(0)$, $x_i$ and $y_i$
which is inverted to yield
$\dot{x}_i(0)$ in terms of $x_i$ and $y_i$. For our purposes it is
sufficient to determine $\dot{x}_i(0)$ to order
$(\Delta t)^{3/2}$. We find
\begin{eqnarray}
\dot{x}_{i} ( t=0 ) & = & \frac{1}{ \Delta t } (x-y)_{i}
-i \r \Bigl\{ \frac{1}{2} F_{ij} (x-y)_j
-  \frac{1}{6} F_{ij,l} (x-y)_j  (x-y)_l \nonumber \\
& + & \frac{1}{24}  F_{ij,kl} (x-y)_j (x-y)_k  (x-y)_l  \nonumber \\
& - &  \frac{1}{12} i \Delta t \r F_{ij}  F_{jk} (x-y)_k + \cdots \Bigr\}.
\label{xexp}
\end{eqnarray}
This result combined with (\ref{lag}) and (\ref{exp}) leads to (\ref{action}).

Factoring out $ \exp[- \frac{1}{\hbar} S_{cl} ]$ from (\ref{res}) we left with
\begin{eqnarray}
&\ &<x|\exp(- \Delta t \hat{H} / \hbar)|y>  =
(2 \pi \hbar \Delta t )^{-n/2} \exp[- \frac{1}{\hbar} S_{cl} ] \nonumber \\
&\ & \exp \Bigl( \frac{i \Delta t }{12} \r [ F_{ki,k} (x-y)_i
- \frac{1}{2} F_{ki,kj} (x-y)_i (x-y)_j ] \nonumber \\
&\ & \hspace{3cm} - \frac{ (\Delta t)^2}{48} \r^2 F^2 \Bigr).
\end{eqnarray}
We expect also a factor of $( \det D_{ij} )^{1/2}$ to be present in the
propagator,where $D_{ij}$ is the Van Vleck matrix
\begin{equation}
D_{ij}(x,y;\Delta t)=- \frac{ \partial }{ \partial x_i}
\frac{ \partial }{ \partial y_j} S_{cl}(x,y; \Delta t ).
\end{equation}
In fact, the remaining terms are just the Van Vleck determinant,
and no other terms are present to order
$ (\Delta t)^2 $. Note that the $F^2$ term which yields the trace anomaly
in two dimensions\cite{vanfio}, is part of the Van Vleck determinant,
whereas a corresponding term $\hbar R$ in curved spacetime is not
contained in the corresponding Van Vleck determinant. This is not
surprising since the $F^2$ is a one-loop effect whereas the $\hbar R$ is
a two loop effect.

Our final result, thus, reads
 \begin{eqnarray}
&\ &<x| \exp(- \Delta t \hat{H} / \hbar)|y>= \nonumber \\
&\ & \hspace{1cm} (2 \pi \hbar )^{-n/2}
(\det D_{ij})^{1/2} \exp[- \frac{1}{\hbar} S_{cl} ]
[1+ O( (\Delta t)^{5/2} )].
\end{eqnarray}

\section{Derivation of the path integral.}

The time-discretized path integral with $N-1$ intermediate steps is given by
\begin{eqnarray}
&\ & <x| \exp (-T\hat{H} / \hbar) |y> = \lim_{N \rightarrow \infty}
\Big( \frac{N}{2 \pi \hbar T} \Big)^{n/2}
\int \prod_{i=1}^n \prod_{\alpha=1}^{N-1}
\Big[ dx_{\alpha i}  \big( \frac{N}{2 \pi \hbar T} \big)^{1/2} \Big]
\nonumber \\
&\ & \exp \Big\{ - \frac{1}{2 \epsilon \hbar} \sum_{\alpha=1}^N
( x_{\alpha-1} - x_\alpha )^2
+ \frac{ie}{\hbar c} \sum_{\alpha=1}^N A_{i}(x_{\alpha-1})
( x_{\alpha-1} - x_\alpha )_i \nonumber \\
&\ & \hspace{1.5cm} - \frac{ie}{2 \hbar c} \sum_{\alpha=1}^N
A_{i,j} (x_{\alpha-1})( x_{\alpha-1} - x_\alpha )_i
( x_{\alpha-1} - x_\alpha )_j \Big\},
\end{eqnarray}
where $\alpha$ is the discretization index and letters from the middle
of the latin alphabet like i, j, k etc. are spacetime indices.
$\epsilon \equiv T/N$ and $x_0 \equiv x$, $x_N \equiv y$.
To obtain this result we inserted $N-1$ complete
sets of states $|x_\alpha><x_\alpha|$ and used the result (\ref{res})
for the matrix element
$<x_{\alpha-1}| \exp (-\epsilon \hat{H} / \hbar) |x_\alpha>$
obtained from the Hamiltonian approach. We kept only the terms up to
order $\epsilon$ (the first three lines in (\ref{res})), because only
these terms will contribute in the limit $N \rightarrow \infty$.
We decompose $x_{\alpha i}$ as
\begin{equation}
x_{\alpha i} = z_{\alpha i} + \xi_{\alpha i}. \label{decomp}
\end{equation}
The $z_{\alpha i}$ yield the classical trajectory of a free particle and
satisfy the equation
\begin{equation}
z_{(\alpha+1) i} - 2 z_{\alpha i} + z_{(\alpha-1) i} = 0,  \label{dieq}
\end{equation}
with boundary conditions
\begin{equation}
z_{0 i} = x_i, z_{N i} = y_i. \label{bound}
\end{equation}
In the limit $N \rightarrow \infty$ (\ref{dieq}) becomes the field equation
of the action for a free particle.
(\ref{dieq}) with the boundary conditions (\ref{bound}) can be solved to
yield
\begin{equation}
z_{\alpha i}= x_i + \frac{\alpha}{N} ( y - x )_i. \label{ddec}
\end{equation}
The $\xi$'s are the quantum fluctuations with boundary conditions
$\xi_{0 i}=\xi_{N i}=0$.
We go over to the mode variables using the transformation
\begin{equation}
\xi_{\alpha i} = \sum_{k=1}^{N-1} y_{i}^k \sin \alpha k \pi / N.  \label{tr}
\end{equation}
The path integral becomes
\begin{eqnarray}
&\ & \exp (-\frac{(x-y)^2}{2 \hbar T} ) \lim_{N \rightarrow \infty}
\Big( \frac{N}{2 \pi \hbar T} \Big)^{n/2}
\int \prod_{i=1}^n \prod_{k=1}^{N-1} \Big[ d y^k_i
\big( \frac{N^2}{4 \pi \hbar T} \big)^{1/2} \Big] \nonumber \\
&\ &\exp \big(-\frac{N^2}{2 \hbar T} \sum_{k=1}^{N-1} (y_{i}^k)^2
( 1 - \cos k \pi / N ) \big)\nonumber \\
&\ & \exp \frac{ie}{\hbar c} \sum_{\alpha=1}^{N} \Big\{
\big[ A_{i}(z_{\alpha-1}) + A_{i,j}(z_{\alpha-1}) \xi_{(\alpha-1) j}
\nonumber \\
&\ & \hspace{1.5cm} + \frac{1}{2} A_{i,jk}(z_{\alpha-1})
\xi_{(\alpha-1) j} \xi_{(\alpha-1) k} + \cdots \big]
\big[ \frac{1}{N} (x-y)_i + ( \xi_{(\alpha-1) i} - \xi_{\alpha i} ) \big]
\nonumber \\
&\ & -\frac{1}{2} \big[ A_{i,j}(z_{\alpha-1})
+ A_{i,jk}(z_{\alpha-1}) \xi_{(\alpha-1) k} \nonumber \\
&\ &\hspace{1.5cm}
+ \frac{1}{2} A_{i,jkl}(z_{\alpha-1})
\xi_{(\alpha-1) k} \xi_{(\alpha-1) l} + \cdots \big]{\rm x} \nonumber \\
&\ & {\rm x}\big[ \frac{1}{N} (x-y)_i
+ ( \xi_{(\alpha-1) i} - \xi_{\alpha i}) \big]
\big[ \frac{1}{N} (x-y)_j + ( \xi_{(\alpha-1) j}
- \xi_{\alpha j}) \big] \Big\}, \label{inte}
\end{eqnarray}
where the $\xi$'s are functions of the modes $y_i^k$ as in (\ref{tr}).
A summation over $i=1, \ldots, n$ is understood in all terms in the exponent.
We have used  that the matrix
$M_{\alpha k}=\sqrt{2/N} \sin \alpha k \pi / N$ is an orthogonal matrix.
This produces the extra factor of $N/2$ in the measure for $y_i^k$.
We rescale the modes
\begin{eqnarray}
v^k_i &=& \Big( \frac{ 2 N^2
( 1- \cos \frac{ k \pi}{ N } )}{k^2 \pi^2 } \Big)^{1/2} y^k_i \nonumber \\
 &\equiv& \lambda(k)^{-1} y^k_{\mu}. \label{resc}
\end{eqnarray}
The kinetic term becomes
\begin{equation}
\exp \big(-\sum_{k=1}^{N-1} \frac{(k \pi)^2}{4 \hbar T} (v^k_i)^2 \big),
\end{equation}
while the measure becomes
\begin{equation}
\Big(\frac{N}{ 2 \pi \hbar T } \Big)^{n/2}
\prod_{i=1}^{n} \prod_{k=1}^{N-1} ( \frac{N^2}{ 4 \pi \hbar T} )^{1/2}
\Big( \frac{k^2 \pi^2}{ 2 N^2 (1-\cos\frac{k \pi}{N})} \Big)^{1/2} dv^k_i.
\end{equation}
This expression can be simplified by using the product formula
\begin{equation}
\prod_{k=1}^{N-1} 2 ( 1- \cos k \pi / N ) = N, \label{for}
\end{equation}
which is a special case ($x\rightarrow1$) of the formula
\begin{equation}
\left[ \prod_{k=1}^{N-1}( x - \cos k \pi / N ) \right]^2 =
\frac{2^{1-2N}}{x^2-1}
{ \rm Re} \big[ -1 + ( x+ i \sqrt{ 1- x^2} )^{2N} \big]. \label{product}
\end{equation}
To derive this formula, one uses that $x^2-1$ times the left hand side is
proportional to
$ \prod_{k=1}^{2N-1} (x-\cos k \pi / N )$.
The measure now becomes
\begin{equation}
( 2 \pi T \hbar )^{ -n/2 } \prod_{i=1}^n \prod_{k=1}^{N-1}
\big( \frac{ \pi k^2 }{ 4 T \hbar } \big)^{1/2} dv^{k}_i.
\end{equation}
Thus, the $N$ dependence of the kinetic term and the measure have
disappeared after the rescaling (the  $N$ appears only in the upper limit
of the sum) and the $N \rightarrow \infty$ limit can be easily taken.
One finds
\begin{equation}
( 2 \pi T \hbar )^{ -n/2 } \prod_{i=1}^n \prod_{k=1}^{\infty}
\big( \frac{ \pi k^2 }{ 4 T \hbar } \big)^{1/2} dv^{k}_i,  \label{measure}
\end{equation}
for the measure and
\begin{equation}
\exp \big(-\frac{1}{2 \hbar T} \int_{-1}^{0} d \tau \dot{\xi}^2 \; \big)=
\exp \big(-\sum_{k=1}^{\infty} \frac{(k \pi)^2}{4 \hbar T} (v^k_i)^2 \big),
\label{kinetic}
\end{equation}
for the kinetic term, where $\xi_i$ is the continuum limit of (\ref{tr})
\begin{equation}
\xi_i(\tau)= \sum_{k=1}^{\infty} v^k_i \sin k \pi \tau. \label{ctr}
\end{equation}
The propagator for the modes obtained from the kinetic term reads
\begin{equation}
\langle v_{i}^{m} v_{j}^{n} \rangle =
\frac{ 2 T \hbar }{ \pi^2 n^2} \delta_{ij} \delta^{mn}. \label{propag}
\end{equation}

At this point we have obtained the measure of the continuum theory,
given in \cite{fio,vanfio}, and the kinetic term and the propagators for
the modes are also that of the continuum theory. It remains to take the
limit in the interaction terms.

\section{The limit $N \rightarrow \infty$ in the interaction terms.}

The interaction terms in (\ref{inte}) can be recast as follows
\begin{eqnarray}
&\ & \exp \frac{ie}{\hbar c} \sum_{\alpha=1}^{N} \Bigg\{
\Big[ A_{i}(z_{\alpha-1}) + A_{i,j}(z_{\alpha-1}) \xi_{(\alpha-1) j}
+ \cdots \Big]
\Big[ \frac{1}{N} (x-y)_i \Big] \nonumber \\
&\ & + A_i(z_{\alpha-1}) (\xi_{(\alpha-1) i} - \xi_{\alpha i}) \nonumber \\
&\ & + A_{i,j}(z_{\alpha-1})
\Big[(\xi_{(\alpha-1) i} - \xi_{\alpha i}) \xi_{(\alpha -1) j} -
\frac{1}{2} (\xi_{(\alpha-1) i} - \xi_{\alpha i})
(\xi_{(\alpha-1) j} - \xi_{\alpha j}) \Big] \nonumber \\
&\ & + \cdots
+ \frac{1}{q!} A_{i,j_1 \cdots j_q}(z_{\alpha-1})
\Big[(\xi_{(\alpha-1) i} - \xi_{\alpha i}) \xi_{(\alpha -1) j_1}
- \nonumber \\
&\ & - \frac{q}{2} (\xi_{(\alpha-1) i} - \xi_{\alpha i})
(\xi_{(\alpha-1) j_1} - \xi_{\alpha j_1}) \Big]
\xi_{(\alpha -1) j_2} \cdots \xi_{(\alpha -1) j_q} + \cdots \Bigg\},
\label{inte1}
\end{eqnarray}
where according to (\ref{tr}) and (\ref{resc}) we have now
\begin{equation}
\xi_{\alpha i} = \sum_{k=1}^{N-1} v_i^k \lambda(k) \sin \alpha k \pi/N.
\label{modes}
\end{equation}
The first line in (\ref{inte1}) is coming solely from the $A_i (x-y)_i$
term in (\ref{inte}), whereas the rest is a combination of both
the $A_i (x-y)_i$ and the $A_{i,j} (x-y)_i (x-y)_j$ terms.
Actually only one part of the latter contributes, namely the
one which is proportional to
$(\xi_{(\alpha-1) i} - \xi_{\alpha i}) (\xi_{(\alpha-1) j} - \xi_{\alpha j})$.
The rest tends to zero when
$N \rightarrow \infty$, as will be clear at the end of this section.
These terms are not shown in (\ref{inte1}).

We now proceed to show that the first line in (\ref{inte1}) limits to
\begin{equation}
\exp \frac{ie}{\hbar c} \int_{-1}^0 d\tau A_i(x(\tau))(x-y)_i, \label{zterm}
\end{equation}
whereas the rest of (\ref{inte1}) limits to
\begin{equation}
\exp \frac{ie}{\hbar c} \int_{-1}^0 d\tau A_i(x(\tau)) \dot{\xi}_i.
\label{xiterm}
\end{equation}
The $x(\tau)$ is the contimuum limit of (\ref{decomp})
\begin{equation}
x_i(\tau) = z_i(\tau) + \xi_i(\tau),
\end{equation}
where  $z_i(\tau)$ is the continuum limit of (\ref{ddec})
\begin{equation}
z_i(\tau) = x_i - \tau (y-x)_i, \label{cdec}
\end{equation}
and $\xi_i(\tau)$ is given in (\ref{ctr}).

There are eight different kinds of terms which we encounter trying to
take the limit $N \rightarrow \infty$ in (\ref{inte1}).
We can have terms with or without $\dot{\xi}$.
Each of them can contain an even or odd number
of quantum fields. (In the latter case only interference terms
can be studied since the expectation value of odd number
of quantum fields is trivially zero). Finally, in each case we
can also have an additional factor of $(\alpha /N)^p$ coming from
the expansion of $A_i(z_{\alpha})$ around $x_i$ (see (\ref{ddec})).
We will illustrate with six examples how the limit $N \rightarrow \infty$
can be rigorously taken in all cases.
The basic idea is that expanding (\ref{inte1}) leads to {\em uniformly}
convergent series for $N$ in the whole interval $1 \leq N < \infty$,
and therefore the limit $N \rightarrow \infty$ can be taken
before the summation over the modes will be performed.

\subsection{Examples with only $\xi$'s.}

The case of only $\xi$'s is relatively easier than the case where
$\dot{\xi}$'s are involved.
This case covers all the terms in the first line in (\ref{inte1}) and also
all the extra terms we would have if we had started with an additional
scalar potential $V(x)$.
We  give two examples where two $\xi$'s are involved. In the first one
the two $\xi$'s are coming from the same $S_{\rm int}$ whereas in  the
second case we deal with an intereference term.
We use the latter case to illustrate how one deals with  factors like
$(\alpha /N)^p$.

\subsubsection{Example 1.}

We shall show that
\begin{equation}
\lim_{N \rightarrow \infty} \langle \frac{1}{N}
\sum_{\alpha=1}^{N-1} \xi_{\alpha i} \xi_{\alpha j} \rangle
=\langle \int_{-1}^0 d\tau \xi_i(\tau) \xi_j(\tau) \rangle,  \label{1ex}
\end{equation}
where `$\langle \; \rangle$' means path integral average.
We start with the left hand side
\begin{equation}
\lim_{N \rightarrow \infty} \frac{1}{N}
\sum_{k,l=1}^{N-1} \langle v_i^k v_j^l \rangle \lambda(k) \lambda(l)
\sum_{\alpha=1}^{N-1} \sin \alpha k \pi/N \sin \alpha l \pi/N.
\end{equation}
The propagator is given in (\ref{propag}) and the $\lambda(k)$  in
(\ref{resc}). Combining the product of the two sines into a sum of two
cosine functions, the summation over $\alpha$ yields
$N/2\; \delta_{kl}$. Hence, the left-hand side of (\ref{1ex}) yields
\begin{equation}
\frac{1}{2} (2 \hbar T) \delta_{ij} \lim_{N \rightarrow \infty}
\sum_{k=1}^{N-1} \frac{1}{4N^2 \sin^2 k \pi/2 N}.
\end{equation}
We remove the $N$-dependence in the summation symbol by extending the sum
to infinity,
rewriting the sum as
\begin{equation}
\sum_{k=1}^{\infty} f_k(N), \label{ser}
\end{equation}
where
\begin{eqnarray}
f_k(N)= \left\{ \begin{array}{ll}
0 & \mbox{if $k>N-1$} \nonumber \\
1/(4N^2 \sin^2 k \pi/2 N) & \mbox{if $k \leq N-1$}.
\end{array}
\right.
\end{eqnarray}
We view $f_k(N)$ as a function of $N$.
Since $k \leq N-1$, clearly $k \pi/2N < \pi/2$. Using the inequality
$2 \theta/ \pi \leq \sin \theta \leq \theta$ for $0 \leq \theta \leq \pi/2$
we get an upper bound for the summands
\begin{equation}
|f_k(N)| \leq \frac{1}{4N^2 (k^2/N^2)} = \frac{1}{4 k^2}. \label{upper1}
\end{equation}
Since the series $\sum_{k=1}^{\infty} (2k)^{-2}$ is convergent, we conclude
that (\ref{ser}) is
uniformly convergent in $N$ for the whole interval $1 \leq N < \infty$.
Thus, we can interchange the limit of $N$ tending  to infinity with the
summation over $k$. Using (\ref{propag}), we obtain
\begin{eqnarray}
\frac{1}{2} (2 \hbar T) \delta_{ij} \sum_{k=1}^{\infty} \frac{1}{k^2 \pi^2}
&=& \sum_{k,l=1}^{\infty} \langle v_i^k v_j^l \rangle
(\frac{1}{2} \delta^{kl}) \nonumber \\
&=& \langle \int_{-1}^0 d\tau \xi_i(\tau) \xi_j(\tau) \rangle.
\end{eqnarray}
This proves (\ref{1ex}).

\subsubsection{Example 2.}

In our second example we will prove that
\begin{equation}
\lim_{N \rightarrow \infty}  \frac{1}{N^2} \langle \sum_{\alpha,\beta=1}^{N}
\big(\frac{\alpha-1}{N}\big) \big(\frac{\beta-1}{N}\big)
\xi_{(\alpha-1) i} \xi_{(\beta-1) j} \rangle =
\langle \int_{-1}^0 d\tau d\tau' \tau \tau' \xi_i(\tau) \xi_j(\tau') \rangle.
 \label{exa1}
\end{equation}
This term is encountered when we expand the term with
$A_{i,j}(z_{\alpha-1}) \xi_{(\alpha-1) j}$ around $x_i$ in the
first line in (\ref{inte1}), and then  use two
$S_{\rm int}$.
We start again with the left hand side
\begin{equation}
\lim_{N \rightarrow \infty} \sum_{k,l=1}^{N-1} \langle v_i^k v_j^l
\rangle \lambda(k) \lambda(l)
\frac{1}{N^2} \sum_{\alpha,\beta=1}^{N-1} \big(\frac{\alpha}{N}\big)
\big(\frac{\beta}{N}\big)
\sin \alpha k \pi /N \sin \beta l \pi /N. \label{ex2}
\end{equation}
The summation over $\alpha$ and $\beta$ can be easily performed by
observing that all the cases with $(\alpha/N)^p$
factors can be obtained from the ones with no such factors by just
introducing  temporarily an extra parameter $r$
in the argument of one of the sines and then differentiating appropriate
number of times.
So, in our case we write
\begin{equation}
\frac{1}{N} \sum_{\alpha=1}^{N-1} \big(\frac{\alpha}{N}\big)
\sin \alpha k \pi /N=
-\frac{1}{k \pi N} \frac{d}{dr} \sum_{\alpha=1}^{N-1}
\cos r \alpha k \pi /N \Big|_{r=1}.
\end{equation}
The summation over $\alpha$ can be easily performed by writing the cosine
as the real part of an exponential.
The result is
\begin{equation}
\frac{1}{N} \sum_{\alpha=1}^{N-1} \big(\frac{\alpha}{N}\big)
\sin \alpha k \pi /N=
-  \frac{(-1)^k}{2N \tan k \pi /2N}. \label{sumex2}
\end{equation}
Using (\ref{sumex2}), (\ref{propag}) and (\ref{resc}), (\ref{ex2}) becomes
\begin{equation}
\lim_{N \rightarrow \infty} \delta_{ij} (2 \hbar T)
\sum_{k=1}^{N-1} \frac{1}{4N^2 \sin^2 k \pi/2N}
\frac{ \cos^2 k \pi/2N }{4N^2 \sin^2 k \pi/2N}.
\end{equation}
Using the same arguments as in the first example we conclude that the
series over $k$ is uniformly convergent. Therefore the limit
$N \rightarrow \infty$ can be taken keeping $k$ fixed.
The result is
\begin{equation}
\sum_{k,l=1}^{\infty} \big(\frac{2 \hbar T}{k^2 \pi^2}
\delta_{ij} \delta^{kl}\big)
\big[\frac{-(-1)^k}{k \pi}\big] \big[\frac{-(-1)^l}{l \pi}\big] =
\langle \int_{-1}^0 d\tau d\tau' \tau \tau' \xi_i(\tau) \xi_j(\tau') \rangle,
\end{equation}
which proves (\ref{exa1}).

The generalization of these two examples to many $\xi$'s is straightforward.
In every case we first reduce the summation of product of sines to the
summations of a single sine or cosine by using the trigonometric formulas
\begin{eqnarray}
\sin a \sin b & = & \frac{1}{2} [ \cos (a-b) - \cos (a+b) ], \label{tr1} \\
\sin a \cos b & = & \frac{1}{2} [ \sin (a+b) + \sin (a-b) ]. \label{tr2}
\end{eqnarray}
Then we use the results of our previous examples.

\subsection{Examples with $\xi$'s and $\dot{\xi}$'s.}

The case where a $\dot{\xi}$ is involved is more complicated. One naively
expects that in the limit
$N \rightarrow \infty$ the sum
$\sum_{\alpha=0}^{N-1} (\xi_{\alpha j_1} - \xi_{(\alpha+1) j_1})
\xi_{\alpha j_2} \cdots \xi_{\alpha j_q}$
becomes
$ (-1)^{q+1} \int d\tau \dot{\xi}_{j_1}(\tau)
\xi_{j_2}(\tau) \cdots \xi_{j_q}(\tau)$.
( The factor $(-1)^{q+1}$ is due to the fact that $\xi_{(\alpha+1) j}$,
corresponds to a $\tau$-value which is smaller than that of $\xi_{\alpha j}$).
Actually this would have been true if we were allowed to take the limit
$N \rightarrow \infty$ inside the summation (for $k$ fixed). To see this we
insert the mode expansion  for $\xi$ into
the sum
\begin{eqnarray}
&\ & \sum_{\alpha=0}^{N-1} (\xi_{\alpha j_1} - \xi_{(\alpha+1) j_1})
\xi_{\alpha j_2} \cdots \xi_{\alpha j_q} =
\sum_{k_1, \cdots, k_q=1}^{N-1} v_{j_1}^{k_1} \cdots v_{j_q}^{k_q}
\lambda(k_1) \cdots \lambda(k_q) \nonumber \\
&\ & \hspace {0.5cm}\Big[ (1 - \cos k_1 \pi/N)
\sum_{\alpha=0}^{N-1} \sin \alpha k_1 \pi/N \cdots
\sin \alpha k_q \pi/N \nonumber \\
&\ & \hspace{0.5cm} - \sin k_1 \pi/N
\sum_{\alpha=0}^{N-1} \cos \alpha k_1 \pi/N \sin \alpha k_2 \pi/N \cdots
\sin \alpha k_q \pi/N \Big]. \label{naive}
\end{eqnarray}
The sums over $\alpha$ is of order $N$. For fixed $k_1$ the factor
$(1 - \cos k_1 \pi/N)$ tends to $1/N^2$, whereas
the $\sin k_1 \pi/N$ goes as $1/N$. Hence,  the first term inside the
square brackets in (\ref{naive}) naively tends to zero for $N$ going to
infinity and the second one gives the correct continuum limit.
However, a more careful analysis shows that this naive limit is not correct.
Consider, for example, the expectation value for  the case $q=2$.
In the term which is naively zero the sum over
$\alpha$ of
$\sin \alpha k_1 \pi/N \sin \alpha k_2 \pi/N$
gives $(N/2) \delta^{k_1 k_2}$, while the propagator combines with the
$\lambda$'s and cancels the factor $(1 - \cos k_1 \pi/N)$.
The final result is that this term has a limit
$(1/4)(2 T \hbar) \delta_{j_1 j_2}$. The same in true for any $q$,
namely both terms have non-vanishing finite limit. Similar results
hold for the terms which were produced by commutators in the Hamiltonian
approach (the last three lines in (\ref{inte})). Naively all these terms
tend to zero for $N$ going to infinity, but careful analysis reveals a
finite result. In fact the terms coming from commutators just cancel the
contribution from the first term in the square brackets in (\ref{naive}),
so that at the end the naive limit gives the correct result!

\subsubsection{Example 3.}

Consider the terms in the third line in (\ref{inte1})
\begin{eqnarray}
&\ & \sum_{\alpha=0}^{N-1}
\big[(\xi_{\alpha i} - \xi_{(\alpha+1) i}) \xi_{\alpha j}
- \frac{1}{2}(\xi_{\alpha i} - \xi_{(\alpha+1) i})
(\xi_{\alpha j} - \xi_{(\alpha+1) j})\big] = \nonumber \\
&\ & \sum_{\alpha=0}^{N-1}(\xi_{\alpha i} - \xi_{(\alpha+1) i})
\frac{(\xi_{\alpha j} + \xi_{(\alpha+1) j})}{2}. \label{deriv}
\end{eqnarray}
We will show that it limits  to
$\int_{-1}^0 d\tau \dot{\xi}_i(\tau) \xi_j(\tau)$.
We insert the mode expansion for the $\xi$'s and  we use the
trigonometric formula for the decomposition of
$\sin (a+1) k \pi/N$.
The terms proportional to $(1 - \cos k \pi/N)$ indeed cancel each other.
One is left with
\begin{eqnarray}
&\ & \frac{1}{2} \sum_{k,l=1}^{N-1} v_i^k v_j^l \lambda(k) \lambda(l)
\big[ \sin l \pi/N \sum_{\alpha=0}^{N-1} \cos \alpha l \pi/N
\sin \alpha k \pi/N \nonumber \\
&\ & \hspace{3cm} - \sin k \pi/N \sum_{\alpha=0}^{N-1}
\cos \alpha k \pi/N \sin \alpha l \pi/N \big]. \label{xi}
\end{eqnarray}
The expectation value of (\ref{xi}) vanishes since the expression within the
square brackets is antisymmetric in $k,l$
whereas the propagator for the modes provides a $\delta^{kl}$.
Thus, it is trivially equal to
$\langle \int_{-1}^0 d\tau \dot{\xi}_i(\tau) \xi_j(\tau) \rangle$
which is also equal to zero.

\subsubsection{Example 4.}

The case with one $\dot{\xi}$ and three $\xi$'s is more delicate.
It corresponds to the case $q=3$ in (\ref{inte1}).
We will prove that
\begin{eqnarray}
&\ & \lim_{N \rightarrow \infty} \frac{1}{3!} A_{i,jkl}(x)
\langle \sum_{\alpha=0}^{N-1}
\Big[(\xi_{\alpha i} - \xi_{(\alpha+1) i}) \xi_{\alpha j} \nonumber \\
&\ & \hspace{2cm} - \frac{3}{2}(\xi_{\alpha i} - \xi_{(\alpha+1) i})
(\xi_{\alpha j} - \xi_{(\alpha+1) j})\Big]
\xi_{\alpha k} \xi_{\alpha l} \rangle  \label{ex3} \\
&\ & = \frac{1}{3!} A_{i,jkl}(x) \sum_{\alpha=0}^{N-1}
 \Big[ \langle (\xi_{\alpha i} - \xi_{(\alpha+1) i})
\frac{(\xi_{\alpha j} + \xi_{(\alpha+1) j})}{2} \rangle
\langle \xi_{\alpha k} \xi_{\alpha l} \rangle \nonumber \\
&\ & \hspace{2cm} + \mbox{cyclic in $j, k, l$} \Big] \label{exa3} \\
&\ &= 0 = \frac{1}{3!} A_{i,jkl}(x)
\langle \int_{-1}^0 d\tau \dot{\xi}_i(\tau) \xi_j(\tau) \xi_k(\tau)
\xi_l(\tau) \rangle
\label{exam3}
\end{eqnarray}
Since the $A_{i,jkl}$ is symmetric in $j, k, l$, we can symmetrize the
second line in (\ref{ex3}).
This yields three terms, each with a factor 1/2.  Applying Wick's theorem
we expect each of them to give three contractions. However, only one
contraction is non-zero, namely
$\langle (\xi_{\alpha } - \xi_{(\alpha+1) })
(\xi_{\alpha } - \xi_{(\alpha+1) }) \rangle
\langle \xi_{\alpha } \xi_{\alpha } \rangle$.
We now show that the other two possible contractions are zero.
Consider the case
\begin{eqnarray}
&\ &\langle (\xi_{\alpha } - \xi_{(\alpha+1) }) \xi_{\alpha } \rangle
\langle (\xi_{\alpha } - \xi_{(\alpha+1) }) \xi_{\alpha } \rangle =
\nonumber \\
&\ & \sum_{k_1,k_2=1}^{N-1}
\big( \frac{2 \hbar T}{2N^2 (1 - \cos k_1 \pi/N)} \big)
\big( \frac{2 \hbar T}{2N^2 (1 - \cos k_2 \pi/N)} \big) \nonumber \\
&\ & \Big[(1 - \cos k_1 \pi/N)(1 - \cos k_2 \pi/N)
\sum_{\alpha=0}^{N-1} \sin^2 \alpha k_1 \pi/N \sin^2 \alpha k_2 \pi/N
\label{1} \\
&\ & - (1 - \cos k_1 \pi/N) \sin k_2 \pi/N \frac{1}{2}
\sum_{\alpha=0}^{N-1} \sin^2 \alpha k_1 \pi/N \sin 2 \alpha k_2 \pi/N
\label{2} \\
&\ & - (1 - \cos k_2 \pi/N) \sin k_1 \pi/N \frac{1}{2}
\sum_{\alpha=0}^{N-1} \sin^2 \alpha k_2 \pi/N \sin 2 \alpha k_1 \pi/N
\label{3} \\
&\ & + \sin k_1 \pi/N \sin k_2 \pi/N \frac{1}{4}
\sum_{\alpha=0}^{N-1} \sin 2 \alpha k_1 \pi/N \sin 2 \alpha k_2 \pi/N \Big],
\label{4}
\end{eqnarray}
where we have supressed the spacetime indices.
The terms (\ref{2}) and (\ref{3}) are clearly zero due to the summation
over $\alpha$.
Furthermore, (\ref{1}) and (\ref{4}) each vanish in  the limit
$N \rightarrow \infty$.
Using this result, Wick's theorem, and the symmetrization in $j, k, l$,
(\ref{ex3}) becomes
\begin{eqnarray}
&\ & \frac{1}{3!} A_{i,jkl}(x) \sum_{\alpha=0}^{N-1}
\Big[ \langle (\xi_{\alpha i} - \xi_{(\alpha+1) i}) \xi_{\alpha j} \rangle
\langle \xi_{\alpha k} \xi_{\alpha l} \rangle - \nonumber \\
&\ &  - \frac{1}{2}
\langle (\xi_{\alpha i} - \xi_{(\alpha+1) i})
(\xi_{\alpha j} - \xi_{(\alpha+1) })\rangle
\langle \xi_{\alpha k} \xi_{\alpha l} \rangle
+ \mbox{(cyclic in $j, k, l$)} \Big].
\end{eqnarray}
This indeed agrees with (\ref{exa3}).
We will show that (\ref{exa3}) is equal to zero.
We substitute  the mode expansion for the $\xi$'s into (\ref{exa3}).
After some trigonometry we get
\begin{eqnarray}
&\ &\langle (\xi_{\alpha} - \xi_{\alpha+1})
\frac{(\xi_{\alpha} + \xi_{\alpha+1})}{2} \rangle
\langle \xi_{\alpha} \xi_{\alpha} \rangle=
\nonumber \\
&\ & \sum_{k_1,k_2=1}^{N-1}
\frac{1}{2} \big( \frac{2 \hbar T}{2N^2 (1 - \cos k_1 \pi/N)} \big)
\big( \frac{2 \hbar T}{2N^2 (1 - \cos k_2 \pi/N)} \big)
\sin^2 \alpha k_2 \pi/N \nonumber \\
&\ & \Big[- \sin^2k_1 \pi /N \cos 2 \alpha k_1 \pi /N
- \frac{1}{2} \sin 2 k_1 \pi/N \sin 2 \alpha k_1 \pi/N \Big]
, \label{error}
\end{eqnarray}
where we have again suppressed  the spacetime indices.
The second term in (\ref{error})
vanishes due to the summation over $\alpha$. The first one
tends to zero in the limit $N \rightarrow \infty$.
Here we use the summation formula
\begin{equation}
\sum_{\alpha=0}^{N-1} \cos \alpha k_1 \pi/N \sin^2 \alpha k_2 \pi/N
= - \frac{N}{4} \delta_{k_1,2k_2}. \label{deltasum}
\end{equation}
To prove this formula we first use trigonometric formulas to reduce
the summation in (\ref{deltasum}) to summations over a single cosine
function and then we perform these summations.
Thus indeed (\ref{exa3}) is equal to zero.
In (\ref{exam3}), combining the cosine with a sine, and combining the
two remaining sine functions, leads to double-angle sine functions whose
integral vanishes. This proves the continuum limit for the term $q=3$ in
(\ref{inte1}).

It is straightforward to generalize to the case of one $\dot{\xi}$ and
arbitrary number of $\xi$'s.
In every case we first use the symmetry of
$A_{i,j_1 \cdots j_q}$ in $j_1, \ldots, j_q$ to symmetrize the
$(\xi_{\alpha} - \xi_{\alpha+1})^2 \xi_{\alpha} \ldots \xi_{\alpha} $
term, so that $q$ terms are obtained. Then Wick's theorem gives $q$
contractions for the
$(\xi_{\alpha} - \xi_{\alpha+1})\xi_{\alpha} \xi_{\alpha} \ldots \xi_{\alpha}$
term, but just one contraction for each of the $q$ terms since all but one
contraction vanish.
The $q$ terms from
$(\xi_{\alpha} - \xi_{\alpha+1})\xi_{\alpha} \xi_{\alpha} \ldots \xi_{\alpha}$
term  combine with the $q$ terms from the symmetrization of
$(\xi_{\alpha} - \xi_{\alpha+1})^2 \xi_{\alpha} \ldots \xi_{\alpha}$
to yield $q$ terms of the form
$(\xi_{\alpha} - \xi_{\alpha+1}) \frac{(\xi_{\alpha} + \xi_{\alpha+1})}{2}
\xi_{\alpha} \ldots \xi_{\alpha}$. Using similar arguments as in the case
of (\ref{exa3}) one can show that the generalization of (\ref{exa3}) also
vanishes and, therefore, is trivially equal to the
continuum case.

\subsubsection{Example 5.}

We now consider examples of interference. The first example concerns with
the interference of two terms, each with an even number of  $\xi$ fields.
We take twice the third line in (\ref{inte1}).
We will show that
\begin{eqnarray}
&\ &\lim_{N \rightarrow \infty}
\sum_{\alpha,\beta=0}^{N-1}
\frac{1}{2} \langle (\xi_{\alpha i_1} - \xi_{(\alpha+1) i_1})
(\xi_{\alpha j_1} + \xi_{(\alpha+1) j_1}) \nonumber \\
&\ & \hspace{2cm} \frac{1}{2} (\xi_{\beta i_2} - \xi_{(\beta+1) i_2})
(\xi_{\beta j_2} + \xi_{(\beta+1) j_2}) \rangle, \label{ex4}
\end{eqnarray}
is equal to
\begin{equation}
\langle \int_{-1}^0 d\tau d\tau' \dot{\xi}_{i_1}(\tau) \xi_{j_1}(\tau)
\dot{\xi}_{i_2}(\tau') \xi_{j_2}(\tau') \rangle. \label{4cxi}
\end{equation}
As we have shown in (\ref{deriv}), the contraction of the first two
(or last two) factors in (\ref{ex4}) vanishes, so that only two
contractions remain. The summations over $\alpha$ and $\beta$ can
be performed (use (\ref{xi}) twice and that (\ref{xi}) vanishes for $k=l$)
to yield
\begin{eqnarray}
&\ &\lim_{N \rightarrow \infty}
\sum_{k_1 \neq l_1; k_2 \neq l_2}
\big[ \langle v_{i_1}^{k_1} v_{i_2}^{k_2} \rangle
\langle v_{j_1}^{l_1} v_{j_2}^{l_2} \rangle
+ \langle v_{i_1}^{k_1} v_{j_2}^{l_2} \rangle
\langle v_{j_1}^{l_1} v_{i_2}^{k_2} \rangle \big] \nonumber \\
&\ &\lambda(k_1) \lambda(l_1) \lambda(k_2) \lambda(l_2)
\frac{1}{4}[1-(-1)^{k_1+l_1}][1-(-1)^{k_2+l_2}] \nonumber \\
&\ &\Big[
\frac{\sin k_1 \pi/N \sin l_1 \pi/N}{\cos k_1 \pi/N - \cos l_1 \pi/N} \Big]
\Big[
\frac{\sin k_2 \pi/N \sin l_2 \pi/N}{\cos k_2 \pi/N - \cos l_2 \pi/N}
\Big]. \label{4xi}
\end{eqnarray}
Each propagator gives a $\delta$-function, so we left with a double sum.
Combining each propagator with the corresponding two factors of $\lambda$,
we get
\begin{eqnarray}
&\ &(2 \hbar T)^2
(\delta_{i_1 i_2} \delta_{j_1 j_2} - \delta_{i_1 j_2} \delta_{j_1 i_2})
\nonumber \\
&\ &\lim_{N \rightarrow \infty} \sum_{k,l=1}^{N-1}
\big(\frac{1}{4N^2 \sin^2 k \pi/2N}\big)
\big(\frac{1}{4N^2 \sin^2 l \pi/2N}\big)
\frac{1}{2}[1-(-1)^{k+l}] \nonumber \\
&\ & \hspace{1.5cm}
\Big[\frac{\sin k \pi/N \sin l \pi/N}{\cos k \pi/N - \cos l \pi/N} \Big]^2
\nonumber \\
&\ & \equiv (2 \hbar T)^2
(\delta_{i_1 i_2} \delta_{j_1 j_2} - \delta_{i_1 j_2} \delta_{j_1 i_2})
\lim_{N \rightarrow \infty} I(N). \label{4xil}
\end{eqnarray}
Using the trigonometric formula for the sine of double angle and the one
which expresses the difference of cosines
as a product of sines we get
\begin{equation}
I(N)= \sum_{k,l=1}^{N-1} \frac{1}{2}[1-(-1)^{k+l}] \frac{1}{4N^4}
\frac{\cos^2 k \pi/2N \cos^2 l \pi/2N}{\sin^2 (l-k) \pi/2N
\sin^2 (k+l) \pi/2N}.
\end{equation}
We split this sum into two sums according to whether $k+l$ is smaller
or larger than $N$
\begin{equation}
I(N)=I_1(k+l \leq N)+I_2(k+l>N),
\end{equation}
where
\begin{equation}
I_1(N)= \sum_{k,l=1}^{\infty}g_{kl}^{(1)}(N),
\end{equation}
and
\begin{eqnarray}
g_{kl}^{(1)}= \left\{ \begin{array}{ll}
0 & \mbox{if $k+l>N$} \nonumber \\
\frac{1}{2}[1-(-1)^{k+l}] \frac{1}{4N^4}
\frac{\cos^2 k \pi/2N \cos^2 l \pi/2N}{\sin^2 (l-k) \pi/2N
\sin^2 (k+l) \pi/2N} & \mbox{if $k+l \leq N$}.
\end{array}
\right.
\end{eqnarray}
In $I_2$ we make the transformation
\begin{equation}
k'=N-k,\; l'=N-l,
\end{equation}
so $k'+l' < N$ and $I_2$ becomes
\begin{equation}
I_2(N)= \sum_{k',l'=1}^{\infty}g_{k'l'}^{(2)}(N),
\end{equation}
where
\begin{eqnarray}
g_{k'l'}^{(2)}= \left\{ \begin{array}{ll}
0 & \mbox{if $k'+l' \geq N$} \nonumber \\
\frac{1}{2}[1-(-1)^{k'+l'}] \frac{1}{4N^4}
\frac{\sin^2 k' \pi/2N \sin^2 l' \pi/2N}{\sin^2 (l'-k') \pi/2N
\sin^2 (k'+l') \pi/2N} & \mbox{if $k'+l' < N$}.
\end{array}
\right.
\end{eqnarray}

We shall now again prove that these series converge uniformly in $N$.
\newline
For $0 < k+l \leq N$, we have the upper bound
\begin{equation}
\sin (k+l) \pi/2N \geq (k+l)/N,
\end{equation}
using the inequality $\sin \theta \geq 2 \theta / \pi$ valid
for $0 \leq \theta \leq \pi /2$. From the same inequality we also get
\begin{equation}
|\sin (l-k) \pi/2N| \geq |l-k|/N,
\end{equation}
since $-N \leq (l-k) \leq N$.
Hence, an upper limit for the summands in $I_1$ can be found which is
independent of $N$
\begin{equation}
|g_{kl}^{(1)}(N)| \leq \frac{1}{4(k^2 - l^2)^2} \frac{1}{2}[1-(-1)^{k+l}].
\end{equation}
The same upper limit holds for $g_{k'l'}^{(2)}(N)$. (From (\ref{xi})
it follows that $g_{kl}^{(1)}$ and $g_{kl}^{(2)}$
vanish at $k=l$).
The double series
$\sum_{k,l=1}^{\infty} [1-(-1)^{k+l}]/[8(k^2 - l^2)^2]$
is convergent. Actually,
apart for the factor 1/8, this is exactly the series we analytically evaluate
in section 5.
Thus, the limit $N \rightarrow \infty$ can be taken keeping fixed $k$ and $l$.
The result is that $I_2(N)$ tends to zero whereas $I_1(N)$ tends to
\begin{equation}
\sum_{k,l=1}^{\infty} [1-(-1)^{k+l}] \frac{2}{\pi^4 (l^2-k^2)^2}.
\end{equation}
Going back to (\ref{4xil}) we get
\begin{equation}
(\delta_{i_1 i_2} \delta_{j_1 j_2} - \delta_{i_1 j_2} \delta_{j_1 i_2})
\sum_{k,l=1}^{\infty}
\big(\frac{2 \hbar T}{k^2 \pi^2}\big) \big(\frac{2 \hbar T}{l^2 \pi^2}\big)
\Big(-[1-(-1)^{k+l}]\frac{kl}{l^2-k^2}\Big)^2. \label{befcon}
\end{equation}
Using
\begin{eqnarray}
\int_{-1}^0 d\tau (k\pi) \cos k \pi \tau \sin l \pi \tau=
\left\{ \begin{array}{ll}
0 & \mbox{if $k=l$} \nonumber \\
-[1-(-1)^{k+l}]\frac{kl}{l^2-k^2} & \mbox{if $k \neq l$},
\end{array}
\right.
\end{eqnarray}
we find that (\ref{befcon}) indeed reproduces (\ref{4cxi}).

\subsubsection{Example 6.}

We now give an example of interference with two terms, each with an odd
number of quantum fields. The basic features are the same, the algebra though
is much more laborious. We take the second term in the first line of
(\ref{inte1}) and the term with $q=2$. We expand $A_{l,n}(z_{\alpha-1})$
around $x_i$.
We will prove that
\begin{eqnarray}
&\ & \lim_{N \rightarrow \infty} \frac{1}{2!}
A_{i,jk}(x) A_{l,mn}(x) (x-y)_l (y-x)_m \langle \frac{1}{N}
\sum_{\alpha, \beta=0}^{N-1}
\Big[(\xi_{\alpha i} - \xi_{(\alpha+1) i}) \xi_{\alpha j} \nonumber \\
&\ & \hspace{2cm}
- \frac{2}{2}(\xi_{\alpha i} - \xi_{(\alpha+1) i})
(\xi_{\alpha j} - \xi_{(\alpha+1) j})\Big]
\xi_{\alpha k} \big(\frac{\beta}{N}\big) \xi_{\beta n} \rangle  \label{ex5} \\
&\ & = - \frac{1}{2!} A_{i,jk}(x) A_{l,mn}(x) (x-y)_l (y-x)_m \nonumber \\
&\ & \hspace{2cm}
\langle \int_{-1}^0 d\tau d\tau' \dot{\xi}_i(\tau) \xi_j(\tau)
\xi_k(\tau) \tau' \xi_n(\tau') \rangle,
\end{eqnarray}
where the relative minus sign is due to the difference in sign between
(\ref{ddec}) and (\ref{cdec}).
Following the same procedure as in the case  with one  $\dot{\xi}$ and odd
number of $\xi$'s we first symmetrize w.r.t $j, k$.
We will study only the contraction
$\langle i\ j \rangle \langle k\ n \rangle$
since the contraction
$\langle i\ k \rangle \langle j\ n \rangle$
is equal to this one and the last one,
$\langle i\ n \rangle \langle j\ k \rangle$,
can be studied in a similar way
, where we abbreviate
the $\xi$'s by their spacetime indices. From now on the factor
$\frac{1}{2!} A_{i,jk}(x) A_{l,mn}(x) (x-y)_l (y-x)_m$
is implied and the spacetime indices are
supressed.
We substitute the mode expansion for the $\xi$'s in (\ref{ex5}).
The summation over $\beta$ is given in (\ref{sumex2}).
After some trigonometry (\ref{ex5}) becomes
\begin{eqnarray}
&\ & \lim_{N \rightarrow \infty} (2 \hbar T)^2
\sum_{k,l=1}^{N-1} \frac{1}{2N^2 (1 - \cos k \pi/N)}
\frac{1}{2N^2 (1 - \cos l \pi/N)} \nonumber \\
&\ & \big[ - \frac{(-1)^l}{2N \tan l \pi/2N} \big] \nonumber \\
&\ & \Big\{ (1 - \cos k \pi/N) \frac{1}{2}
\sum_{\alpha=0}^{N-1} (1 - \cos 2 \alpha k \pi/N) \sin \alpha l \pi/N
\label{10} \\
&\ & - \sin k \pi/N \big( \frac{N}{4} \delta^{2k,l} \big) \label{11} \\
&\ & - \frac{1}{2} (1 - \cos k \pi/N)^2 \frac{1}{2}
\sum_{\alpha=0}^{N-1} (1 - \cos 2 \alpha k \pi/N) \sin \alpha l \pi/N
\label{12} \\
&\ & - \frac{1}{2} (1 - \cos k \pi/N)(1 - \cos l \pi/N) \nonumber \\
&\ & \hspace{3cm} \frac{1}{2} \sum_{\alpha=0}^{N-1}
(1 - \cos 2 \alpha k \pi/N) \sin \alpha l \pi/N \label{13} \\
&\ & + (1 - \cos k \pi/N) \sin k \pi/N \big( \frac{N}{4} \delta^{2k,l} \big)
\label{14} \\
&\ & + \frac{1}{2} (1 - \cos k \pi/N) \sin l \pi/N
\big( -\frac{N}{4} \delta^{2k,l} \big) \label{15} \\
&\ & + \frac{1}{2} (1 - \cos l \pi/N) \sin k \pi/N
\big( \frac{N}{4} \delta^{2k,l} \big) \label{16} \\
&\ & - \frac{1}{2} \sin^2 k \pi/N \frac{1}{2}
\sum_{\alpha=0}^{N-1} (1 + \cos 2 \alpha k \pi/N)
\sin \alpha l \pi/N \label{17} \\
&\ & - \frac{1}{2} \sin k \pi/N \sin l \pi/N
\frac{1}{2} \sum_{\alpha=0}^{N-1} \sin 2 \alpha k \pi/N
\cos \alpha l \pi/N \Big\}. \label{18}
\end{eqnarray}
The terms (\ref{10}) and (\ref{11}) are coming from the
$(\xi_{\alpha} - \xi_{\alpha+1}) \xi_{\alpha}$ term.
The former is cancelled by the $(\xi_{\alpha} - \xi_{\alpha+1})^2$
term and the latter gives the continuum limit.
Indeed, the term (\ref{10}) is cancelled exactly by the terms
(\ref{12}), (\ref{13}), (\ref{17}) and (\ref{18}).
The terms (\ref{14}), (\ref{15}) and (\ref{16}) vanish in the limit
$N \rightarrow \infty$.
It remains to take the limit $N \rightarrow \infty$ in (\ref{11}).
The term (\ref{11}) can be rewritten as
\begin{equation}
- (2 \hbar T)^2
\sum_{k=1}^{N-1} \frac{1}{128 N^4}
\frac{ \cos k \pi/N}{ \sin^2 k \pi/2N \sin^2 k \pi/N}. \label{lastt}
\end{equation}
We split the sum in two sums, the first running from 1 to $N/2 - 1$
and the second from $N/2$ to $N-1$.
In the first one an upper bound for the summands can be found by
using the same inequalities as in the first example,
\begin{equation}
\Big|\frac{1}{128 N^4}
\frac{ \cos k \pi/N}{ \sin^2 k \pi/2N \sin^2 k \pi/N}
\Big| \leq \frac{1}{512 k^4}.
\end{equation}
In the second sum we make the transformation $k' = N - k$. The series becomes
\begin{equation}
\sum_{k'=1}^{N/2}
\frac{1}{128 N^4} \frac{ \cos k' \pi/N}{ \cos^2 k' \pi/2N \sin^2 k' \pi/N} =
\sum_{k'=1}^{N/2}
\frac{1}{32 N^4} \frac{ \sin^2 k' \pi/2N  \cos k' \pi/N}{\sin^4 k' \pi/N}.
\end{equation}
An upper bound for the summands of this series is given by
\begin{equation}
\Big|
\frac{1}{32 N^4}
\frac{ \sin^2 k' \pi/2N \cos k' \pi/N}{ \sin^4 k' \pi/N}
\Big|
\leq \frac{1}{512 k'^4}.
\end{equation}
Therefore the series (\ref{lastt}) is uniformly convergent.
So the limit $N \rightarrow \infty$ in (\ref{11}) can be performed
before the summation.
The result reads
\begin{eqnarray}
&\ & - (2 \hbar T)^2 \sum_{k,l=1}^{\infty}
\frac{1}{k^2 \pi^2} \frac{1}{l^2 \pi^2}
\Big[- \frac{(-1)^l}{l \pi} \Big]
(k \pi) \big( \frac{1}{4} \delta^{2k,l} \big) \nonumber \\
&\ &= - \int_{-1}^0 d\tau d\tau'
\langle \dot{\xi}_i(\tau) \xi_j(\tau) \rangle
\tau'  \langle \xi_k(\tau) \xi_n(\tau') \rangle
\end{eqnarray},
which is what we wanted to prove.

One can easily check now that the terms of (\ref{inte}) which were
omitted in (\ref{inte1}) indeed tend to zero.
These terms are those in the last three lines in (\ref{inte}) except
for the terms proportional to
$(\xi_{(\alpha-1) i} - \xi_{\alpha i}) (\xi_{(\alpha-1) j} - \xi_{\alpha j})$.
All of them are equal to $1/N$ times terms which were proven finite
in the limit $N \rightarrow \infty$.

Combining (\ref{kinetic}), (\ref{zterm}) and (\ref{xiterm}) we get the
continuum action $S_{\rm config}$
\begin{equation}
S_{\rm config} = \frac{1}{T} \int_{-1}^{0} d \tau
[ \frac{1}{2} \dot{x}_{i} \dot{x}_{i}
-i \r T \dot{x}_{i} A_{i} ]
=\int_{-T}^{0} dt
[ \frac{1}{2} \dot{x}_{i} \dot{x}_{i}
-i \r \dot{x}_{i} A_{i} ]. \label{coaction}
\end{equation},
where in the last step we have rescaled the the time $\tau=t/T$.

\section{Evaluation of the path integral.}

In the continuum path integral with action (\ref{coaction}) we set
$T=\Delta t$ and then we evaluate it to order $(\Delta t)^2$.
The derivation of the path integral indicates how to evaluate it.
First we decompose $x_i(\tau)$ into a function $z_i( \tau )$ and
a quantum part $\xi_i( \tau )$
\begin{equation}
x_i ( \tau ) = z_{i} ( \tau ) + \xi_{i} ( \tau ).
\end{equation}
The function $z_i( \tau )$ is not a solution of the classical field
equations, but rather of the field equations
corresponding to $L_0 = \dot{x}^2 /2 $. It satisfies the same boundary
conditions as $ x_i ( \tau ) $ and hence
is given by
\begin{equation}
z_{i} = x_i - \tau ( y-x )_i.
\end{equation}
It follows that the quantum field vanishes at the boundary
\begin{equation}
\xi_{i} ( \tau = 0 ) = \xi_{i} ( \tau = -1 ) = 0.
\end{equation}
Since the eigenfunctions of $S_0$ with these boundary conditions are the
functions $ \sin (n \pi \tau) $, we expand the quantum field on a
trigonometric basis \cite{dekker,fio,vanfio}
\begin{equation}
\xi_{i} = \sum_{n=1}^{ \infty } v_{i}^{n} \sin (n \pi \tau ).
\end{equation}
The propagator for the modes is obtained by using only the part quadratic
in velocities and reads
\begin{equation}
\langle v_{i}^{m} v_{j}^{n} \rangle =
\frac{ 2 \Delta t \hbar }{ \pi^2 n^2} \delta_{ij} \delta^{mn},
\end{equation}
as follows from the measure in (\ref{measure}).
If we would multiply this result with two sine functions and sum over
$m$ and $n$, we would recover the result of \cite{vanfio,bas} for
$\langle \xi_i(\tau_1) \xi_j(\tau_2) \rangle$. However, we shall work
here entirely in terms of modes.

The $S_{\rm int}$, up to the order we are interested in, is given by
\begin{eqnarray}
S_{\rm int} & = & \frac{i}{ \hbar } \r  \int_{-1}^{0} \!d \tau \,
\big[ (x-y)_i + \dot{ \xi }_i \big] \big\{ A_{i}( z( \tau ) )
+ A_{i,j}( z( \tau ) ) \xi_j \nonumber \\
&\ & + \frac{1}{2} A_{i,jk}( z( \tau ) ) \xi_j \xi_k
+ \frac{1}{3!} A_{i,jkl}( z( \tau ) ) \xi_j \xi_k \xi_l + \cdots  \big\}.
\label{int}
\end{eqnarray}
We factor out all the terms which do not depend on quantum fields
\begin{eqnarray}
\exp \Big[&-& \frac{1}{2 \hbar \Delta t} (x-y)^2
+ \frac{i}{\hbar} \r \Big\{ A_i (x-y)_i
- \frac{1}{2} A_{i,j} (x-y)_i (x-y)_j \nonumber \\
& + & \frac{1}{3!} A_{i,jk} (x-y)_i (x-y)_j (x-y)_k \nonumber \\
& - &  \frac{1}{4!} A_{i,jkl} (x-y)_i (x-y)_j (x-y)_k (x-y)_l \Big\} \Big].
\label{noq}
\end{eqnarray}
Observe that (\ref{noq})  differs from  $ \exp(- S_{\rm cl}/ \hbar )$
by just one term (namely the $F^2$ term in
(\ref{action}) ). We will recover this missing term from a tree graph
(see (\ref{last})).
The reason for the absence of the $F^2$ term from (\ref{noq}) is that
$z_i( \tau )$ does not satisfy the full field equations but rather the
field equations of $L_0 = \dot{x}^2 /2 $.

Using only one factor of $S_{\rm int}$ we get the following contribution
\begin{eqnarray}
(\frac{ie}{ \hbar c}) \! \int_{-1}^{0} \!d \tau \,
& \big[ & A_{i,j} ( z( \tau) )
\langle \dot{ \xi}_i( \tau ) \xi_j ( \tau ) \rangle \nonumber \\
&+&
\frac{1}{2} A_{i,jk}( z( \tau) ) \langle \xi_j( \tau ) \xi_k( \tau ) \rangle
(x-y)_i \nonumber \\
&+&
\frac{1}{3!} A_{i,jkl}( z( \tau) )
\langle \dot{ \xi}_i( \tau ) \xi_j( \tau ) \xi_k( \tau ) \xi_l( \tau )
\rangle \big]. \label{one}
\end{eqnarray}
The first two terms are one loop contributions and are of order
$ \Delta t $ and higher (since $(x-y)_i$ is of order
$(\Delta t)^{1/2})$, while the last one is a
2-loop contribution of order $(\Delta t)^2$ and higher. However, performing
the $ \tau $-integration the 2-loop contribution of order $(\Delta t)^2$
vanishes because combining the sine and cosine functions one always ends
up with a sine of a double angle whose integral vanishes.
The first term is a superficially divergent tadpole, but using mode
regularization (i.e., first evaluate the integrals for a finite number
of modes, and then let the number of modes tend to infinity) one finds
that it is, in fact, finite.
This is thus a property of our regularization scheme, similar to the
property of dimensional regularization which puts equal to zero all
divergences which are not logarithmic divergences.
The first two terms in (\ref{one}) yield
\begin{equation}
i \frac{ \Delta t}{12} \r \big[ F_{ki,k} (x-y)_i
- \frac{1}{2}  F_{ki,kj} (x-y)_i (x-y)_j \big]. \label{v1}
\end{equation}
To get this result we used the known sum $ \zeta (2) =
\sum_{0}^{ \infty } n^{-2} = \pi^2 /6 $.

Two factors of $S_{\rm int}$ yield
\begin{eqnarray}
- \frac{1}{2} ( \frac{e}{ \hbar c} )^2 \int_{-1}^0 \! d \tau d \tau' \!
&\big\{&\![ A_{i,j}(z(\tau)) A_{k,l}(z(\tau'))
\langle \xi_j(\tau) \xi_l(\tau') \rangle (x-y)_i (x-y)_k \nonumber \\
&+&
A_i(z(\tau)) A_{k,l}(z(\tau'))
\langle \dot{\xi}_i(\tau) \xi_l(\tau') \rangle (x-y)_k \nonumber \\
&+&
A_{i,j}(z(\tau)) A_k(z(\tau'))
\langle \xi_j(\tau) \dot{\xi}_k(\tau') \rangle (x-y)_i \nonumber \\
&+& A_i(z(\tau)) A_k(z(\tau'))
\langle \dot{\xi}_i(\tau) \dot{\xi}_k(\tau') \rangle ] \nonumber \\
&+& A_{i,j}(z(\tau))A_{k,l}(z(\tau'))
\langle\dot{\xi}_i(\tau)\xi_j(\tau)
\dot{\xi}_k(\tau')\xi_l(\tau')\rangle\big\}, \label{2s}
\end{eqnarray}
where we have omitted terms which yield zero after the $\tau$-integration
or are of higher order in $\Delta t$.
The first four terms inside the square brackets are tree graphs and combine
to give
\begin{equation}
\frac{4}{ \pi^4} ( \frac{ \Delta t}{ \hbar} ) \r^2
\sum_{k=0}^{ \infty } \frac{1}{ (2k+1)^4 }
F_{ik} F_{kj} (x-y)_i (x-y)_j.  \label{last}
\end{equation}
The sum which appears in (\ref{last}) is  known and is equal to
$\lambda(4)=( 1- 2^{-4} )\zeta (4)= \pi^4 /96$.
Using this result we identify (\ref{last}) as the missing term of
the classical action.
The last term in (\ref{2s}) is a one-loop graph and gives
\begin{equation}
- \frac{2}{ \pi^4} \r^2 F^2 (\Delta t)^2
\sum_{m,n=1; m \neq n}^{ \infty }
\frac{ 1- (-1)^{m+n} }{ (m^2-n^2)^2 }. \label{v2}
\end{equation}
The double sum which appears in (\ref{v2}) seems not tabulated.
Here we give an analytic evaluation of it.
The idea is to extend the limits of summation to $ \pm \infty $
so that linear shifts of the summation variable
are allowed. This can be done by observing that the summand is
symmetric under $ n \rightarrow -n,
m \rightarrow -m, m \leftrightarrow n$. Notice also that only
$m+n={\rm odd}$ contributes.The sum becomes
\begin{eqnarray}
&\ & \sum_{k,l= - \infty , l \neq 0 }^{\infty}
\frac{1}{ [(2k+1)^2 - (2l)^2 ]^2 } = \nonumber \\
&\ & \sum_{k,l= - \infty }^{\infty}
\frac{1}{ [(2k+1)^2 - (2l)^2 ]^2 } - \pi^4 / 48.
\end{eqnarray}
The last double sum can be rewritten by substituting
$2k = 2l+p-1$ for $p$ odd
\begin{eqnarray}
&\ & \sum_{l=- \infty}^{ \infty }
\sum_{p=1,odd}^{ \infty } \frac{1}{p^2( 4l \pm p )^2} = \nonumber \\
&\ & 2 ( \sum_{p=1,odd}^{ \infty } p^{-2} )^2 =
2 ( \lambda(2) )^2 = \pi^4 /32.
\end{eqnarray}
Hence, the sum is equal to $ \pi^4 / 96 $.

(\ref{v2}) together with (\ref{v1}) give the Van Vleck determinant.
The final result is that the path integral correctly reproduces the
propagator found from the Hamiltonian operator approach.
There are further one-loop diagrams which give contribution of higher
order than $ (\Delta t)^2 $. For example, taking twice the term
$ (x-y)_i A_{i,jk} \xi_j \xi_k $ we get a one-loop result proportional
to $ (\Delta t)^3 $.

\section{Conclusions.}

We have proven the 1-1 correspondence between the Hamiltonian approach
and path integration for Hamiltonians of the form
$ \hat{H}=\hat{p}^2 + a^i(\hat{x}) \hat{p}_i + b(\hat{x})$.
The correspondence we found is this: casting the Hamiltonian into the form
\begin{equation}
\hat{H} = \frac{1}{2}
\Big( \hat{p}_i - (\frac{e}{c}) A_i(\hat{x}) \Big)
\Big( \hat{p}^i - (\frac{e}{c}) A^i(\hat{x}) \Big)
+ V(\hat{x}),
\end{equation}
the action to be used in the path integral is
\begin{equation}
S_{\rm config} = \int_{-T}^0 dt
\big[ \frac{1}{2} \delta_{ij} \dot{x}^i \dot{x}^j
- i(\frac{e}{c}) \dot{x}^i A_i(x)  + V(x) \big]. \label{langr}
\label{coraction}
\end{equation}
This result holds for any Hamiltonian, whether it is covariant or not.
The path integral is perturbatively evaluated by treating the term
$\dot{x}^2 /2$ as the free part $S_0$, decomposing $x(\tau)$ into a
background part $z(\tau)$ and a quantum part $\xi(\tau)$, and expanding
$\xi(\tau)$ in terms of eigenfunctions of $S_0$. The measure in the path
integral as well as the action $S_0$ determines the world line propagator
(see (\ref{propag})). (This is, in fact, the only place where the measure
plays a role for us). The other two terms in (\ref{coraction}) yield the
vertices, and one can now evaluate (as we did) the path integral in a
perfectly straightforward and standard manner (``Feynman graphs'').

Of course the expansion of $\xi_i(\tau)$ into modes is well-known
\begin{equation}
\xi_i(\tau)= \sum_{k=1}^{\infty} v^k_i \sin k \pi \tau.
\end{equation}
What we have shown is that all arbitrariness (such as the overall
normalization of the path integral) can be fixed by starting with the
Hamiltonian approach. Moreover, we have given an elementary (though at times
somewhat tedious) proof that the $N \rightarrow \infty$ limit of the
time-discretized path integral exists as far as perturbation theory is
concerned, and indeed yields the continuum path integral with its measure.

The actual proof that the limit $N \rightarrow \infty$ exists was given by
carefully analyzing six examples which cover all cases one encounters in
the perturbative evaluation of the path integral. In each example we found
upper bounds for the infinite series which showed that these series are
uniformly convergent as a function of $N$. This allowed us to take the
limit $N \rightarrow \infty$ inside the summation symbols (i.e., at fixed
mode index $k$).

Our results confirm the lore about path integrals that the naive
$N \rightarrow \infty$ limit in the discretized path integral yields
the correct continuum path integral. However, this came about by an
interesting ``conspiracy'': the higher order terms in the Hamiltonian
evaluation of $<x| \exp (- \Delta t \hat{H} / \hbar) |y>$ which are due
to expanding the exponent and taking into account the commutators between
$\hat{x}$ and $\hat{p}$ operators, {\em cancel} against the terms in the
time-discretized action which seem to (but do not) vanish in the
$N \rightarrow \infty$ limit. Due to this conspiracy it is, after all,
correct to use (\ref{la}) and (\ref{linapp}), omitting all commutators,
to obtain the action to be used in the path integral. Namely, this yields
$h(x,p)=\frac{1}{2} p^2 - \frac{e}{c} A^i(x) p_i
+ \frac{1}{2} (\frac{e}{c})^2 A^2(x) + V(x)$,
and after integrating out the momenta, the naive
$N \rightarrow \infty$ limit yields the correct action
$S_{\rm config}$ in (\ref{coraction}).
However, (\ref{la}) and (\ref{linapp}) by themselves do not yield
the correct propagator.

Our results immediately generalize to quantum field theories with
derivative interactions, such as Yang-Mills theory with gauge fixing
term and ghost action. The Hamiltonian for the gauge and ghost fields
contain again terms linear in momenta. For example, in the Lorentz gauge,
the Hamiltonian reads
\begin{eqnarray}
{\cal H} (gauge) & = &
\frac{1}{2} p(A^a_k)^2 - \frac{1}{2} p(A^a_0)^2
+ p(A^a_k) \partial_k A_0^a \nonumber \\
& + & p(A^a_0) \partial^k A_k^a
+ \frac{1}{4} (G^a_{kl})^2 - g p(A^a_k) f^a_{\ bc}  A_0^b A_k^c,
\end{eqnarray}
and
\begin{equation}
{\cal H} (ghost) = p(b_a) p(c^a) + p(c^a) g f^a_{\ bc} A_0^b c^c
+ (\partial_k b_a) (D_k c^a),
\end{equation}
where $b_a (c^a)$ are the antighost (ghost) fields. Following the results
of this paper one can find
the 1-1 correspondence between operator Hamiltonians and path integral
actions. In particular, one may determine
the operator ordering of ${\cal H} (gauge)$ and ${\cal H} (ghost)$
which corresponds to the usual BRST invariant quantum action in the
configuration space path integral. However, again, the linear
approximation in the Hamiltonian approach yields incorrect results
if one uses it to compute the propagator in the Hamiltonian approach.

Another extension of our results would be to consider phase space path
integrals. In the discretized action we first integrated at some point
over the momenta, and then studied the limit $N \rightarrow \infty$.
One might leave the discretized momenta in the action, and consider the
limit $N \rightarrow \infty$ with momenta present. The continuum action
is expected to be $p \dot{x} - H(p,x)$, i.e., the Legendre transform of
the classical Lagrangian in (\ref{langr}). Again one could introduce
classical trajectories for $p$ and $x$ satisfying $x(0)=x$, $x(T)=y$
and satisfying the Hamilton equations of motion of a suitable
Hamiltonian $H_0$ contained $H$. (This will, of course, fix $p(0)$
and $p(T)$ as well). The quantum deviations $\xi(\tau)$ and $\pi(\tau)$
vanish then at the boundaries and can be expanded into a complete set,
for example again $\sin k \pi t/T$. The measure is expected to come out
unity (except for the factor $(2 \pi \hbar T)^{-1/2}$ mentioned in the
introduction) and propagators and vertices would then be defined.
The problem would be to prove that the limit $N \rightarrow \infty$ of
the discretized theory indeed produce this continuum theory.

We are interested in extending our results to models in curved spacetime
(nonlinear sigma models). This is a well-known problem to which partial
answers have been given in \cite{dewitt} and \cite{bas}. In the propagator
to order $\Delta t$ one finds a term proportional to Ricci tensor $R_{ij}$
contracted with $(x^i - y^i)(x^j - y^j)$, which cannot be written as the
action of a {\em local} functional. Thus, it is not immediately clear what
the continuum action is, and which terms in the limit
$N \rightarrow \infty$ will cancel. However, one can still exponentiate
this term and obtain the discretized action. The $R_{ij}$ term should
become an $R$ term in the continuum theory since at the perturbative level
$(x^i - y^i)(x^j - y^j)$ should be equivalent to $g^{ij} \Delta t$.
However, this would only produce a factor $R/6$ into the action whereas
one needs a factor $R/8$\cite{dekker,fio,dewitt}. Further cancellations
of type studied in section 4 should then indeed reduce $R/6$ to  $R/8$.
Note that this analysis might be done without the need of using
Einstein invariance to go to Riemann normal coordinates, and hence
problems with time-ordering in arbitrary coordinates\cite{dewitt}
would be avoided.
\newline
\\
{\bf Acknowledgments:} We thank Bas Peeters and Jan de Boer for numerous
discussions. Eduard Br\'{e}zin and Jan Pierre Zuber told us that they
had also discovered that it is incorrect to use the linear approximation
to obtain the propagator (see (\ref{la}), (\ref{linapp})).
We hope that our solution as based on the ``conspiracy'' described in
the conclusions will satisfy them and others.
\newline
\\
This work was supported by NSF grant 92-11367.


\begin{thebibliography}{99}
\bibitem{schul} See, for example, L. Schulman,
`Techniques and Applications of Path Integration',
John Wiley and Sons, New York, 1981.
\bibitem{A_W} L. Alvarez-Gaum\'{e} and E. Witten,
Nucl. Phys. {\bf B234}, (1984) 269.
\bibitem{diaz} A. Diaz, W. Troost, P. van Nieuwenhuizen and A. van Proeyen,
Int. J. Mod. Phys. {\bf A4} (1989) 3959;
M. Hatsuda, P. van Nieuwenhuizen, W. Troost and A. van Proeyen,
Nucl. Phys. {\bf B335}, (1990) 166.
\bibitem{dekker} H. Dekker, Physica {\bf 103A} (1980) 586.
\bibitem{fio} F. Bastianelli, Nucl. Phys. {\bf B376}, (1992) 113.
\bibitem{vanfio} F. Bastianelli and P. van Nieuwenhuizen,
Nucl. Phys. {\bf B389}, (1993) 53.
\bibitem{graham} R. Graham, Z. Phys. {\bf B26}, (1977) 281.
\bibitem{book} F. Langouche, D. Roekaerts, and E. Tirapegui,
`Functional Integration and Semeclassical Expansions',
D. Reidel Publishing Company, Dordrecht, Holland, 1982.
\bibitem{bas} B. Peeters, P. van Nieuwenhuizen, ITP-SB-93-51.
\bibitem{vleck} J. Van Vleck, Proc. Natl. Acad. Sci. {\bf 24}, (1928) 178;
C. Morette, Phys. Rev. {\bf 81} (1951) 848.
\bibitem{witt} B. DeWitt, Rev. Mod. Phys. {\bf 29}, (1957) 377.
\bibitem{dewitt} B. DeWitt, `Supermanifolds', $2{nd}$ ed.,
Campridge University Press, 1992.
\end{thebibliography}
\end{document}